\documentclass[3p, 11pt]{elsarticle}
\makeatletter
\def\ps@pprintTitle{%
 \let\@oddhead\@empty
 \let\@evenhead\@empty
 \def\@oddfoot{\centerline{\thepage}}%
 \let\@evenfoot\@oddfoot}
\makeatother
\usepackage{amsfonts,amssymb,amsmath,amsopn,float}
\usepackage{xifthen}
\usepackage{graphicx}
\usepackage{tikz}
\usepackage{pgfplots}
\usepgfplotslibrary{units}
\usepackage{epstopdf}
\usepackage{url}
\usepackage{subcaption}
\usepackage{soul}
\usepackage[plainpages=false,naturalnames,colorlinks=true,linkcolor=blue,bookmarks]{hyperref}
\usepackage[titletoc,title]{appendix} 

\usepackage{fancyhdr}
\usepackage{placeins} 

\usepackage{color}

\newcommand{\bbR}{\mathbb{R}}

\newcommand{\dyad}{\mathbf{\otimes}}
\newcommand{\bzero}{\mathbf{0}}

\newcommand{\bolds}[1]{\boldsymbol{#1}}
\newcommand{\ba}{\bolds{a}}
\newcommand{\bb}{\bolds{b}}

\newcommand{\be}{\bolds{e}}
\newcommand{\bff}{\bolds{f}}
\newcommand{\bg}{\bolds{g}}

\newcommand{\bu}{\bolds{u}}
\newcommand{\bv}{\bolds{v}}

\newcommand{\bx}{\bolds{x}}
\newcommand{\by}{\bolds{y}}
\newcommand{\bz}{\bolds{z}}

\newcommand{\bF}{\bolds{F}}

\newcommand{\bI}{\bolds{I}}

\newcommand{\bT}{\bolds{T}}

\newcommand{\rmd}{{\, \rm d}}

\newcommand{\inffnr}{J\left(\frac{r}{\epsilon}\right)}

\newcommand{\sautoref}[2]{\hyperref[#2]{#1 \ref*{#2}}}

\newdefinition{rmk}{Remark}
\newproof{proof}{Proof}

\usepackage{verbatim}

\usepackage{xspace}

\newcommand*{\ie}{i.e.\@\xspace}

\makeatletter
\newcommand*{\etc}{%
    \@ifnextchar{.}%
        {etc}%
        {etc.\@\xspace}%
}
\makeatother

\begin{document}

\begin{frontmatter}

\title{Peridynamics-based discrete element method (PeriDEM) model of granular systems involving breakage of arbitrarily shaped particles}

\author[aa]{Prashant K. Jha\corref{cor1}\fnref{fn1}}
\ead{pjha@utexas.edu}

\author[bb]{Prathamesh S. Desai\fnref{fn2}}
\ead{pdesai@rice.edu}

\author[cc]{Debdeep Bhattacharya\fnref{fn3}}
\ead{debdeepbh@lsu.edu}

\author[cc]{Robert Lipton\fnref{fn4}}
\ead{lipton@lsu.edu}

\address[aa]{Oden Institute for Computational Engineering and Sciences, The University of Texas at Austin, TX}

\address[bb]{Department of Mechanical Engineering, Rice University, Houston, TX}

\address[cc]{Department of Mathematics, Louisiana State University, Baton Rouge, LA}

\cortext[cor1]{Corresponding author (pjha@utexas.edu)}


\begin{abstract}
  Usage, manipulation, transport, delivery, and mixing of granular or particulate media, comprised of spherical or polyhedral particles, is commonly encountered in industrial sectors of construction (cement and rock fragments), pharmaceutics (tablets), and transportation (ballast). Elucidating particulate media's behavior in concert with particle attrition (i.e., particle wear and subsequent particle fragmentation) is essential for predicting the performance and increasing the efficiency of engineering systems using such media. Discrete element method (DEM) based techniques can describe the interaction between particles but cannot model intra-particle deformation, especially intra-particle fracture. On the other hand, peridynamics provides the means to account for intra-particle deformation and fracture due to contact forces between particles. The present study proposes a hybrid model referred to as \textit{PeriDEM} that combines the advantages of peridynamics and DEM. The model parameters can be tuned to achieve desired DEM contact forces, damping effects, and intra-particle stiffness. Two particle impacts and compressive behavior of multi-particle systems are thoroughly investigated. The model can account for any arbitrarily shaped particle in general. Spherical, hexagonal, and non-convex particle shapes are simulated in the present study. The effect of mesh resolution on intra-particle peridynamics is explicitly studied. The proposed hybrid model opens a new avenue to explore the complicated interactions encountered in discrete particle dynamics that involve the formation of force chains, particle interlocking, particle attrition, wear, and the eventual breakage.
\end{abstract}


\end{frontmatter}

\section{Introduction}
Granular media consists of a collection of mesoscale to macro-scale solid particles. Modeling granular media is a challenging problem as it involves the modeling of contact forces between particles with an arbitrary boundary and deformation of individual particles with fracture/plasticity/corrosion effects. Additional issues beyond contact forces and intra-particle fracture or damage, such as the effect of entrained gas/fluid on granular media dynamics, temperature dependent mechanical properties, could also become important. The discrete element method (DEM) introduced in \citet{cundall1979discrete} provides a framework in which linear and angular displacements and velocities of individual particles are solved using Newton's second law. The contact forces, moments, friction forces, and damping forces between two particles in direct contact are postulated to assume that the particles are spherical (disks in 2D). Particles in DEM are assumed to retain their shape, allowing one to only focus on contact-related interactions. DEM has been applied to problems involving powder dynamics in additive manufacturing \citet{desai2019rheometry}, particle packing, mixing and segregation \citet{labra2009high, yan2016investigating}, particle transport and particle-fluid interaction \citet{feng2007coupled, zhu2007discrete}, among other applications. Nonlocal bonded DEM has also been used to study rock mechanics \citet{desai2017tribosurface}. DEM-based methods have been extended to handle the arbitrarily shaped particles and particle breakage using cohesive interactions \citet{neveu2016fracture, nguyen2017cohesive}. 

Under the high loading, the particle's internal deformation becomes a significant factor in particulate media dynamics. Particles may undergo a change in shape, and they may yield or eventually break. Intra-particle deformation requires modeling each particle as a continuum solid with appropriate constitutive law, e.g., linear/nonlinear elasticity, LEFM, plasticity, \etc. As the particle deforms, it changes its shape and can break into smaller particles; hence the contact forces with adjacent particles become challenging to model. In this study, we consider the peridynamics model for the deformation of individual particles; this enables us to incorporate intra-particle fracture/damage effects naturally. In peridynamics, the internal force at a material point is expressed using a summation of pairwise forces with material points inside a neighborhood of interaction. This contrasts with the classical continuum mechanics in which the divergence of stress (for example, linear elastic mechanics) gives the internal force at a material point. Peridynamics, proposed initially in \citet{silling2000reformulation}, has evolved further with application to key fracture problems \citet{silling2010crack, foster2011energy, bobaru2012meaning, lipton2016cohesive, lipton2019complex, jha2020kinetic}. Because of its ability to naturally incorporate the fracture process and making it easy to couple the fracture process with other physics, the theory has seen great success in modeling fracture in solids \citet{silling2005meshfree, silling2005peridynamic, ha2010studies, diehl2016numerical, behzadinasab2018peridynamics, lipton2018free, jha2020kinetic, wu2020validation}, corrosion \citet{chen2015peridynamic, jafarzadeh2019peridynamic}, erosion \citet{zhang2018coupled}, and porous flow using peridynamics \citet{katiyar2014peridynamic, ouchi2015fully}. A critical feature of peridynamics is that both an elastic deformation and crack emerge from the dynamics without needing additional rule or hypothesis on the motion of the crack tip \citet{silling2000reformulation}. Recently it is shown that the Linear Elastic Fracture Mechanics (LEFM) kinetic relation for crack tip velocity is recovered from the peridynamics equation of motion as the length scale of non-locality approaches zero \citet{lipton2020plane, jha2020kinetic}. 

In this work, we combine the Peridynamics theory with DEM to expand the scope of DEM to a large class of problems where particle deformation and breakage can cause significant changes in particle dynamics. The earlier work \citet{behzadinasab2018peridynamics} applied peridynamics to model both the inter-particle and intra-particle interactions. In \citet{zhu2019modeling} authors proposed a model that utilizes peridynamics for the particle deformation; however, their method significantly differs from \citet{behzadinasab2018peridynamics} and in our work concerning how the contact is applied and how the peridynamics is utilized. This work's major contribution, in contrast to prior results, is the systematic development of a high-fidelity model that handles both inter-particle and intra-particle interactions; the resulting model is intuitive and free from any ad-hoc techniques. In the proposed model, the contact between neighboring particles is governed by a DEM-type contact law describing the normal contact force, frictional force, and damping force. As opposed to the standard DEM, the contact in the proposed model is applied between two particles' material points. Therefore, the particle sees the local boundary of the neighboring particles. In the numerical implementation, the contact acts on a pair of nodes of meshless (also referred to as meshfree) discretization of two bodies. The contact system is activated only when the nodes of opposing bodies are sufficiently close (contact radius distance).

Motivated from the original DEM, the contact forces between two nodes are based on the general spring-dashpot system -- the normal contact force and the damping force are due to the spring stretch and dashpot, respectively. We introduce an additional component in the spring-dashpot system to symbolize the friction force. The spring-dashpot system can be calibrated to achieve the desired magnitude of the normal contact force, friction force, and damping effects. This type of calibration is quite similar to DEM-based methods where the virtual simulator of a granular media is first calibrated \citet{desai2019rheometry, asmar2002validation} using the experimental data and then applied for realistic predictions. In this work, we apply our model to a ladder of problems of increasing complexity. First, we show the damping effect on two-particle collision and demonstrate that the damping parameter can be tuned for the desired damping effect. We consider the impact of mesh size on the two-particle system. Since the contact laws are directly applied in the discretization, some mesh size influence is expected. The results show that mesh effects are within a reasonable range and can be adequately understood; however, we remark that the mesh effect studies in this work are preliminary, and in the future, this will be looked at in more detail. We next showcase the particle damage due to the high-velocity impact of particles. We consider particles of different material properties, in particular the varying fracture strength. We show that when both particles are of the same strength, both sustain damage after impact. When one of the particles has sufficiently high strength, it does not sustain any damage (i.e., the deformation is purely elastic). To exhibit that the model is not restricted by the particle shape and can be easily applied to any arbitrarily shaped particles, we repeat some of the two-particle and fracture tests using hexagon-shaped and non-convex particles. Finally, we apply the model to a compressive test consisting of a collection of particles (500+ particles) inside a rectangular box with the box's top wall moving into the suspension at a prescribed speed. Here, we consider circular and hexagon-shaped particles of varying sizes to study the particulate media's compressive strength. The compressive test reveals the multi-particle system's complex behavior; the media at low loads behaves elastically, and as the loading increases, it begins to yield, leading to complete failure.

While the method shows promising results for the breakage of particulate media, it is computationally costly compared to the traditional DEM-based methods. The computational expense can be attributed to the nonlocal force calculations in each particle and nonlocal search for nodes within two particles expected to collide. With the use of the Kd-tree (k-dimensional tree) routine in the Point Cloud Library (PCL) \citet{Rusu_ICRA2011_PCL, muja2009fast}, we have been able to reduce the computational cost drastically; in the final section, we look at the computational cost of the individual components in more details and discuss several approaches that may reduce the cost further. To promote the development of the model and its application, we have open-sourced the PeriDEM library in this link: \url{https://github.com/prashjha/PeriDEM}. 
This implementation is based on the pre-release version of the NLMech library \citet{jha2019numerical, diehl2020asynchronous} and relies on the HPX \citet{kaiser2020hpx} for the multi-threading computation. 

The paper is organized as follows: In \autoref{s:intro}, we present peridynamics for the intra-particle interaction and a DEM-like model for the inter-particle interaction. 
In \autoref{s:peridemNumericImp}, we discuss the numerical implementation of the proposed model. In \autoref{s:numeric}, we apply the model to various settings. We first analyze the damping effects under the simple two-particle system in \autoref{ss:twop}. Next, we study the effects of mesh size on the inter-particle contact in \autoref{ss:meshEffects}. In \autoref{ss:twoPWall}, we show the high-velocity impact between two particles and resulting fracture. We repeat some tests in prior sections using non-circular particles in \autoref{ss:noncircular}. Having tested the model for a two-particle setting, we apply the model to study the compressive strength of a multi-particle system \autoref{ss:compressiveTest}. In \autoref{s:conclusion}, we present the discussion of the current work, highlight few challenges, and provide future directions.

\section{Development of the PeriDEM model}\label{s:intro}
Let $\varOmega\in \bbR^d$ denote the particulate media domain where $d=2$ or $3$ is the dimension. The media $\varOmega$ consists of particles $\varOmega_{p_i}$, $i=1,2, ...,N$, and is subjected to external forces or displacements altering the configuration of particles within it. Two types of interactions are present in the media: 1) intra-particle interaction in which each particle reacts to the surrounding boundary conditions causing the particle to deform and produce internal forces, and 2) the inter-particle interaction governing the contact between two particles and exchange for forces at the interface. For the first, we consider the peridynamics description of solid deformation. For the second, we propose a DEM-like model. Since the contact is applied between the material points sufficiently close, the model naturally handles the arbitrarily shaped particles.

\subsection{Intra-particle interaction: Peridynamics}
\begin{figure}[!htb] 
  \centering
  \includegraphics[width=0.3\textwidth]{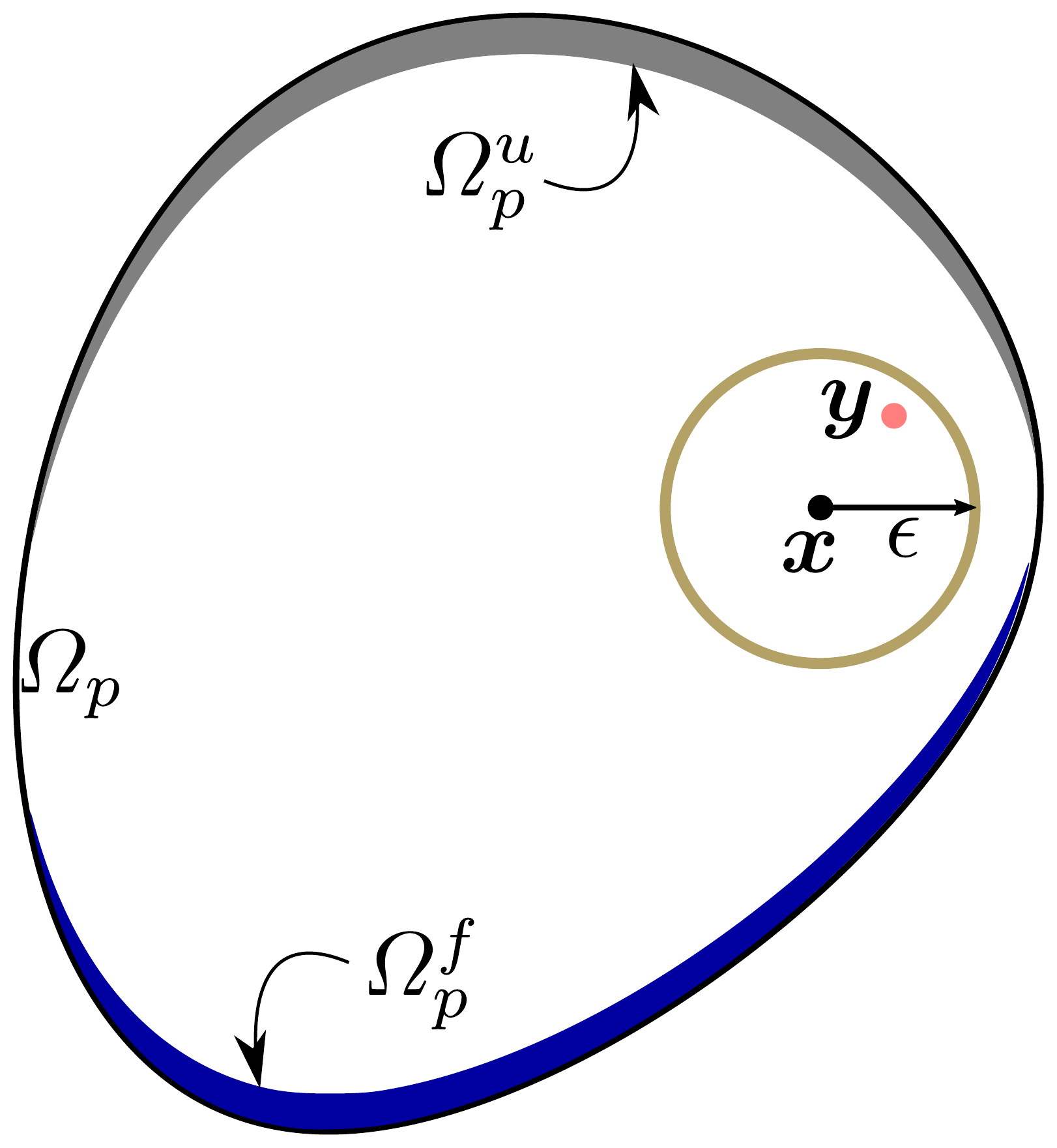}
  \caption{Typical particle domain $\varOmega_p$. The internal force at $\bx$ is due to the interaction of $\bx$ with all the points $\by \in B_\epsilon(\bx)$. The displacement and force boundary conditions are specified in $\varOmega_p^u$ and $\varOmega_p^f$, respectively.}\label{fig:pddomain}  
\end{figure}

Consider a typical particle $\varOmega_p$. Let $\bx\in \varOmega_p$ denote the coordinates of the material point and let $\bu : \varOmega_p \times [0,T] \to \bbR^d$ and $\bv : \varOmega_p \times [0,T] \to \bbR^d$ denote the displacement and velocity fields. At time $t\in [0,T]$, the new coordinates of the material point $\bx\in \varOmega_p$ is given by $\bz(\bx,t) = \bx + \bu(\bx,t)$. In peridynamics, the force at a material point is a result of the pairwise forces acting on the point due to the neighboring points. In a general form, the force at $\bx \in \varOmega_p$ is given by
\begin{align}\label{eq:pd_force}
  \bF(\bx,t; \bu) = \int_{B_\epsilon(\bx)\cap \varOmega_p} \bff(\by, \bx) \rmd\by \,,  
\end{align}
where $B_\epsilon(\bx)$ is the ball of radius $\epsilon$ centered at $\bx$, $\bff(\by,\bx)$  pairwise force acting on material point $\bx$ due to the interaction of $\bx$ with $\by$, and  $\epsilon>0$ the nonlocal length-scale. The motion of points in $\varOmega_p$, and therefore the deformation of $\varOmega_p$, is given by the Newton's second law of motion:
\begin{align}\label{eq:pdmotion}
  \rho \ddot{\bu}(\bx,t) = \bF(\bx, t; \bu) + \bF_{ext}(\bx,t), \qquad \forall\, (\bx,t) \in \varOmega_p \times [0,T]\,,
\end{align}
where $\bF_{ext}$ is the external force such as contact force acting on $\bx$. 
We close the above system by specifying the initial conditions
\begin{align}\label{eq:ic}
	\bu(\bx, 0) = \bu_0(\bx), \qquad \bv(\bx, 0) = \bv_0(\bx), \qquad \forall\, \bx \in \varOmega_p
\end{align}
and the boundary conditions
\begin{align}\label{eq:bc}
\bu(\bx,t) = \bg(\bx), \qquad \forall\, (\bx,t) \in \varOmega_p^u \times [0,T]\,,
\end{align}
where $\bg$ is the prescribed displacement field,  $\varOmega_p^u$ the part of $\varOmega_p$ over which the displacement is prescribed, see \autoref{fig:pddomain}. Under the suitable assumptions on $\bff$, body forces, and initial conditions, it can be shown that the nonlocal dynamics \autoref{eq:pdmotion} converges to the classical continuum mechanics equations $\rho \ddot{\bu}(\bx,t) =\nabla \cdot \bolds{\sigma}$ away from the crack and the nonlocal dynamics delivers the classic equation of crack tip motion $G_c=\mathcal{J}$ where $\mathcal{J}$ is the elastic energy flowing into the tip; see \citet{lipton2016cohesive, jha2019numerical,jha2020kinetic}. Here, $\bolds{\sigma}$ is the Cauchy's stress. 

In the next section, we show the specific form of force $\bff$ used in this work. 

\subsubsection{State-based peridynamics model}
Within the peridynamics formulation there are two classes of models: bond-based and state-based. In the bond-based models, the force between pair of material points only depends on the displacement of the points. In contrast, in the state-based models, the force may also depend on the volumetric deformation of points. In this work, we consider the state-based model \citet{silling2007peridynamic}. Existing literature \citet{silling2007peridynamic, warren2009non} use the concept of {\it states} to describe the model. For simplicity, we present the model without using the concept of {\it states} but using the familiar notion of functions. The force acting on $\bx$ due to $\by$ has the following form \citet{silling2007peridynamic}
\begin{align}\label{eq:bond_via_state_based}
\bff(\by, \bx) = \bT_{\bx}(\by) - \bT_{\by}(\bx)\,,
\end{align}
and therefore the total force density at $\bx$ is
\begin{align}\label{eq:pdStates}
\bF(\bx,t; \bu) = \int_{B_\epsilon(\bx)\cap \varOmega_p} (\bT_{\bx}(\by) - \bT_{\by}(\bx)) \rmd\by\,.
\end{align} 
Fix $r = r(\by, \bx) = |\by - \bx|$. Function $\bT_{\bx}: B_\epsilon(\bx) \to \bbR^d$ defined for each material point $\bx$ is given by \cite{behzadinasab2018peridynamics}
\begin{align}\label{eq:state_Tx}
  \bT_{\bx}(\by) = \inffnr\left[ \kappa\frac{3 r \theta_{\bx}}{m_{\bx}} + G \frac{15 e^d_{\bx}(\by)}{m_{\bx}} \right] \frac{\bz(\by) - \bz(\bx)}{|\bz(\by) - \bz(\bx)|}\,,
\end{align}
where $J$, $J(a) = 1 - a$ when $a \in [0,1]$ and $J(a) = 0$ when $a> 1$, is the influence function. $\kappa$ and $G$ are bulk and shear moduli, $m_{\bx}$ weighted volume of a material point, $\theta_{\bx}$ dilation of material point, and $e^d_{\bx}(\by)$ the deviatoric part of the extension defined for $\by$ in the neighborhood of a material point $\bx$. $m_{\bx}$, $\theta_{\bx}$, and $e^d_{\bx}(\by)$ are given by
\begin{align}\label{eq:mx_thetax}
  m_{\bx} &= \int_{B_\epsilon(\bx)} r^2 \inffnr \rmd\by, \notag \\
  \theta_{\bx} &= \frac{3}{m_{\bx}} \int_{B_\epsilon(\bx)} (|\bz(\by) - \bz(\bx)| - r) r \inffnr \rmd\by, \notag \\
  e^d_{\bx}(\by) &= |\bz(\by) - \bz(\bx)| - r - \frac{r\theta_{\bx}}{3}\,,
\end{align}
We can write $\bT_{\bx}(\by)$ as follows
\begin{align}\label{eq:state_Tx2}
\bT_{\bx}(\by) = \inffnr \left[ r \theta_{\bx} \left(\frac{3\kappa}{m_{\bx}} - \frac{15 G}{3m_{\bx}} \right) + (|\bz(\by) - \bz(\bx)| - r)\left(\frac{15 G}{m_{\bx}}\right)\right] \frac{\bz(\by) - \bz(\bx)}{|\bz(\by) - \bz(\bx)|}\,.
\end{align}

\paragraph{Modeling fracture}
In peridynamics, fracture is incorporated at the bond-level; the bond if stretched beyond a critical stretch is considered broken, \citet{silling2000reformulation}. The crack/fracture is a result of collection of broken bonds. Given the critical energy release rate, the critical stretch $s_0$ beyond which bond is broken is given by \citet{zhu2019peridynamic},
\begin{align}\label{eq:s0}
s_0 = \sqrt{\frac{G_c}{(3\mu + (3/4)^4[\kappa - (5\mu/3)])\epsilon}} \,.
\end{align}
Let $h: \bbR \to [0,1]$ be the function such that 
\begin{align}
	h(s) = \begin{cases}
		1 \qquad \text{ if } s < s_0\,, \\
		0 \qquad \text{ otherwise}\,.
	\end{cases}
\end{align}
Following the implementation in Peridigm library \citet{littlewood2013peridigm}, we modify the $\bT_{\bx}(\by)$ to take into account the bond-breakage as follows:
\begin{align}\label{eq:state_Tx3}
\bT_{\bx}(\by) &= h(s(\by, \bx), t)\inffnr\left[ r \theta_{\bx} \left(\frac{3\kappa}{m_{\bx}} - \frac{15 G}{3m_{\bx}} \right) \right.\notag \\
&\qquad \left. + (|\bz(\by) - \bz(\bx)| - r)\left(\frac{15 G}{m_{\bx}}\right) \right] \frac{\bz(\by) - \bz(\bx)}{|\bz(\by) - \bz(\bx)|} \,,
\end{align}
where recall that $r = |\by - \bx|$. $\theta_{\bx}$ is also modified to account for the damage as follows:
\begin{align}\label{eq:thetax}
\theta_{\bx} &= \frac{3}{m_{\bx}} \int_{B_\epsilon(\bx)} h(s(\by, \bx),t) (|\bz(\by) - \bz(\bx)| - r) r \inffnr \rmd\by\,.
\end{align}

\paragraph{Damage at material points}
We define the damage at point $\bx\in \varOmega_p$ as follows \citet{lipton2019complex}
\begin{align}\label{eq:damage}
  Z(\bx) = \sup_{\by \in B_\epsilon(\bx) \cap \varOmega_p} \frac{|\bu(\by) - \bu(\bx)|}{|\by - \bx|} \frac{1}{s_0}\,,
\end{align}
where $s_0$ is the critical bond-strain. $Z(\bx)<1$ implies that the deformation at the point $\bx$ is elastic and there are no broken bonds in the neighborhood of $\bx$. Whereas,  $Z(\bx)\geq1$ implies that there is atleast one broken bond in the neighborhood of $\bx$. The fracture zone is the region in $\varOmega_p$ consisting of points with one or more broken bonds in the neighborhood, \ie,
\begin{align}\label{eq:damageZone}
  FZ(\varOmega_p) = \{\bx\in \varOmega_p: Z(\bx) \geq 1 \}\,.
\end{align}

\subsection{Inter-particle interaction: DEM-like contact laws}\label{ss:interParticleDEM}
Let $\{\bx_i, V_i\}_{i=1}^n$ are the pair of nodal coordinates and nodal volumes in a meshless discretization of the particle $\Omega_p$. We discuss the meshless discretization in more details in \autoref{ss:peridemImp}. Suppose $\Omega_{p}, \Omega_{p'}$ are the two particles in contact. Traditional DEM-based methods apply contact force at the centroid of particles. In this work, we follow the alternative approach. The idea is to simulate the contact on the discretization nodes of the two bodies when they get sufficiently close. In the proposed model, we consider all three major components of the contact forces – normal force, damping force, and frictional force. We assume that a general spring-dashpot system connects the points of opposing bodies in the contact region. The necessary contact forces result from the deformation of this spring-dashpot system as the points move. Since the contact forces are defined on the pair of discretization nodes, the model is not limited to spherical particles and can be applied to model the contact between bodies of any arbitrary shapes.  Further, the parameters can be tuned to get the desired damping effect, contact strength, and desired aggregate behavior of particle systems. 

Let $\bx\in \Omega_{p}, \bx'\in \Omega_{p'}$ are the discretized nodes of particles $\Omega_{p}, \Omega_{p'}$. Let $V,V'$ denote the volume represented by the nodes $\bx, \bx'$. We denote the current position of $\bx,\bx'$ by $\bz, \bz'$. 
The nodes $\bx, \bx'$ interact only when $|\bz - \bz'| < R_c$, where $R_c$ is the radius of contact. The contact radius is typically chosen as $0.95h$ where $h$ is the mesh size defined as the minimum distance between any two different nodes. We describe the contact forces next.

\begin{figure}[!htb] 
  \centering
  \includegraphics[width=0.5\textwidth]{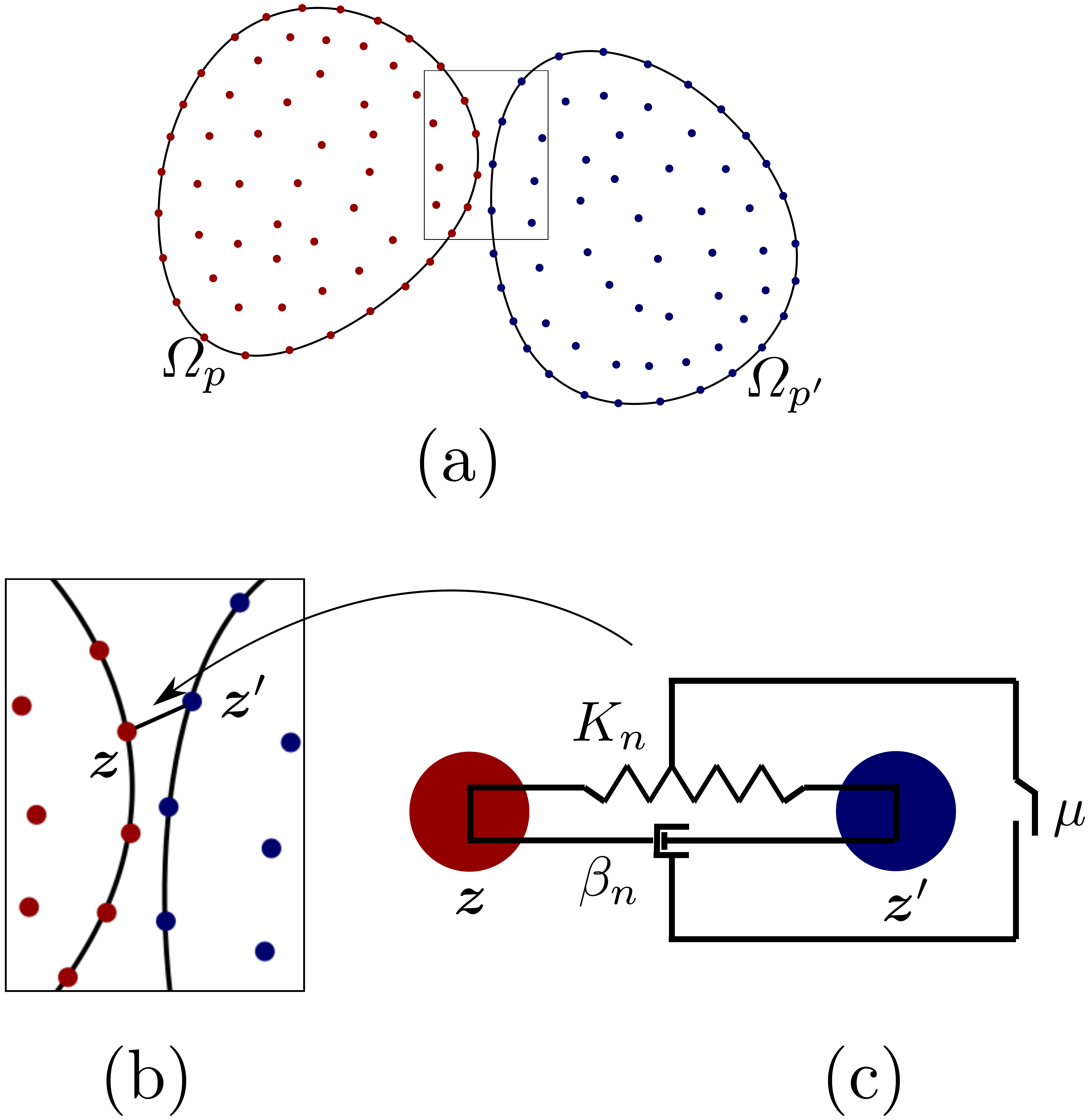}
  \caption{Schematics of the contact forces on two particles $\varOmega_{p}$ and $\varOmega_{p'}$ in contact. In {\bf (a)}, we show the typical discretization of particles; red and blue points are discrete nodes of particles 1 and 2, respectively. In {\bf (b)}, we expand the contact region and show two typical points $\bz,\bz’$ of opposing bodies. In {\bf (c)}, the general spring-dashpot system between the two points is drawn. Recall that $\bx,\bx’$ give the reference coordinates and $\bz, \bz’$ the current coordinates. As points $\bz,\bz’$ move, the spring in {\bf (c)} stretches, producing the reaction force. Since the normal contact force is compressive, the spring only responds when it is compressed compared to its natural length. The relative velocity between $\bz,\bz’$ results in damping effects from the dashpot; the damping force is proportional to the impact velocity between $\bz,\bz’$. Finally, we introduce the switch component (near $\mu$), which symbolizes the frictional force between $\bz, \bz’$. }\label{fig:discretecontact}  
\end{figure}

\subsubsection{Normal contact force}
We assume that the points $\bx, \bx'$ at current position $\bz, \bz'$ are connected by a linear spring with the following properties:
\begin{itemize}
	\item $K_n$ is the spring stiffness constant,
	\item $R_c$ is the equilibrium length (contact radius),
	\item The spring does not resist the tensile loading.
\end{itemize}
Let $\delta$ is the spring stretch defined as the change in the length of spring relative to the equilibrium length, \ie,
\begin{align}
\delta(\bz, \bz') = |\bz  - \bz'| - R_c\,.
\end{align}
Then the normal contact force density on $\bx$ due to $\bx'$ is given by
\begin{align}\label{eq:normalForce}
	\bF_n(\bx', \bx) = \begin{cases}
		K_n \delta(\bz, \bz') V' \be_{n}, &\qquad \text{ if } \delta(\bz, \bz') < 0\,, \\
		\bzero &\qquad \text{ otherwise}\,,
	\end{cases}
\end{align}
where $\be_{n}$ is the unit vector pointing at $\bz'$ from $\bz$, \ie,
\begin{align}
	 \be_n = \frac{\bz' - \bz}{|\bz' - \bz|}\,.
\end{align}
The spring modulus $K_n$ can be related to the bulk moduli of the bodies in contact, see \autoref{ss:peridemImp}. 

\subsubsection{Frictional force}
The friction force on the contacting particles act on a plane with normal $\be_n$. The direction of the force is given by 
\begin{align}\label{eq:tangentPlane}
	\be_t  = \left[ \bI - \be_n \dyad \be_n \right]  \frac{\bv'- \bv}{|\bv'- \bv|}\,,
\end{align}
where $\frac{\bv'- \bv}{|\bv'- \bv|}$ is the unit vector along the relative velocity of impacting nodes, $\ba \dyad \bb$ the matrix product of the two vectors $\ba, \bb$, and $\bI$ the identity tensor. Here, $\bv = \bv(\bx,t)$ and $\bv' = \bv(\bx',t)$ are the velocities of points $\bx, \bx'$. The friction force on $\bx$ due to $\bx'$ is given by (Coulomb's law)
\begin{align}
\bF_t(\bx', \bx) = -\mu |\bF_n(\bx', \bx)| \be_t\,.
\end{align}

\subsection{Damping force}
For damping, we consider two models. In the first model, damping force is defined similar to the normal contact force. In the second model, the damping force is applied between the centroid of the two contacting particles. 

\subsubsection{Damping force between material points}
In addition to spring between point $\bx, \bx'$, we now suppose there is a dashpot with the following properties:
\begin{itemize}
	\item $\beta_n$ is the viscosity of the dashpot,
	\item Damping force is linear with the rate of change in the spring length $\dot{\delta}$ defined as
		\begin{align}
			\dot{\delta} = \frac{d}{dt} \delta(\bz, \bz') = (\bv' - \bv) \cdot \frac{\bz' - \bz}{|\bz' - \bz|}\,,
		\end{align}
	\item Damping force is zero when $\delta(\bz, \bz')> 0$\,. 
\end{itemize}
The damping force density on $\bx$ is then given by
\begin{align}\label{eq:dampingForce}
	\bF_d(\bx', \bx) = \begin{cases}
		\frac{1}{V}\beta_n \dot{\delta}(\bz, \bz') \be_{n}, &\qquad \text{ if } \dot{\delta} (\bz, \bz') < 0\, \text{ and } \delta(\bz, \bz') < 0\,, \\
		\bzero &\qquad \text{ otherwise}\,.
	\end{cases}
\end{align}
The viscosity parameter $\beta_n$ is based on the empirical formula (see \citet{desai2017tribosurface, desai2019rheometry})
\begin{align}\label{eq:viscosity}
	\beta_n = -2C\log(\varepsilon_n) \sqrt{ \frac{\kappa_{eff} R_c m_{eq}}{\pi^2 + \log(\varepsilon_n)^2} }\,,
\end{align}
where $m_{eq}$ is the Harmonic mean of the mass of two nodes in contact, \ie,
\begin{align}
	m_{eq} = \frac{2\rho V \rho' V'}{\rho V + \rho' V'}\,,
\end{align}
where we recall that $V, V'$ are nodal volumes and $\rho, \rho'$ mass density of two nodes in contact. $\kappa_{eff}$ is the effective bulk modulus computed using
\begin{align}\label{eq:effk}
	\kappa_{eff} = \frac{2\kappa_1 \kappa_2}{\kappa_1 + \kappa_2}\,.
\end{align}
$C>0$ is a constant and $\varepsilon_n \leq 1$ is the damping parameter controlling the strength of damping.

\subsubsection{Damping force between particle centers}\label{ss:damping2}
Alternatively, we can apply the damping between the particle centers. Suppose $\bx_c $ and $\bx'_c$ are the centers of particles in contact, $\bz_c, \bz'_c$ their current positions, and $\bv_c, \bv'_c$ their velocities. The distance ${\rm dist}(\varOmega_{p}, \varOmega_{p'})$ between $\varOmega_p, \varOmega_{p'}$ is defined as
\begin{equation*}
    {\rm dist}(\varOmega_{p}, \varOmega_{p'}) = \inf \{|\bz - \bz'|: \bz\in \varOmega_p, \bz'\in \varOmega_{p'}\} \,.
\end{equation*}
We model the damping using the dashpot between $\bx_c$ and $\bx'_c$ with the following properties:
\begin{itemize}
	\item $\bar{\beta}_n$ is the viscosity of the dashpot,
	\item Damping force is linear with the rate of change in the length $\dot{\delta}_c$ defined as
	\begin{align}
		\dot{\delta}_c = (\bv'_c - \bv_c) \cdot \frac{\bz'_c - \bz_c}{|\bz'_c - \bz_c|}\,,
	\end{align}
	\item Damping force acts only when the distance, ${\rm dist}(\varOmega_{p}, \varOmega_{p'})$, between particles is less than the contact radius $R_c$. 
\end{itemize}
The total damping force density at the center of particle $\Omega_{p}$ due to $\Omega_{p'}$ in this case is given by
\begin{align}\label{eq:totalDampingForce}
\bar{\bF}_{d}(\bx'_c, \bx_c) = \begin{cases}
\frac{1}{|\varOmega_p|}\bar{\beta}_n \dot{\delta}_c \frac{\bz'_c - \bz_c}{|\bz'_c - \bz_c|}, &\qquad \text{ if } \dot{\delta}_c < 0 \text{ and } {\rm dist}(\varOmega_{p}, \varOmega_{p'}) < R_c\,, \\
\bzero &\qquad \text{ otherwise}\,.
\end{cases}
\end{align}
The parameter $\bar{\beta}_n$ similar to $\beta_n$ is given by
\begin{align}\label{eq:viscosityBar}
\bar{\beta}_n = -2\bar{C}\log(\bar{\varepsilon}_n) \sqrt{ \frac{\kappa_{eff} R_c M_{eq}}{\pi^2 + \log(\bar{\varepsilon}_n)^2} }\,,
\end{align}
where $M_{eq}$ is the Harmonic mean of the mass of two particles in contact, \ie,
\begin{align}
M_{eq} = \frac{2\rho |\varOmega_{p}| \rho' |\varOmega_{p'}|}{\rho |\varOmega_{p}|  + \rho' |\varOmega_{p'}| }\,.
\end{align}
Here $|\varOmega_{p}|$ denotes the volume (area in 2d) of the domain. $\kappa_{eff}$ is the effective bulk modulus defined in \autoref{eq:effk}, $\bar{C}>0$ a constant, and $\bar{\varepsilon}_n \leq 1$ the damping parameter controlling the strength of damping.

\begin{figure}[!htb] 
	\centering
	\includegraphics[width=0.4\textwidth]{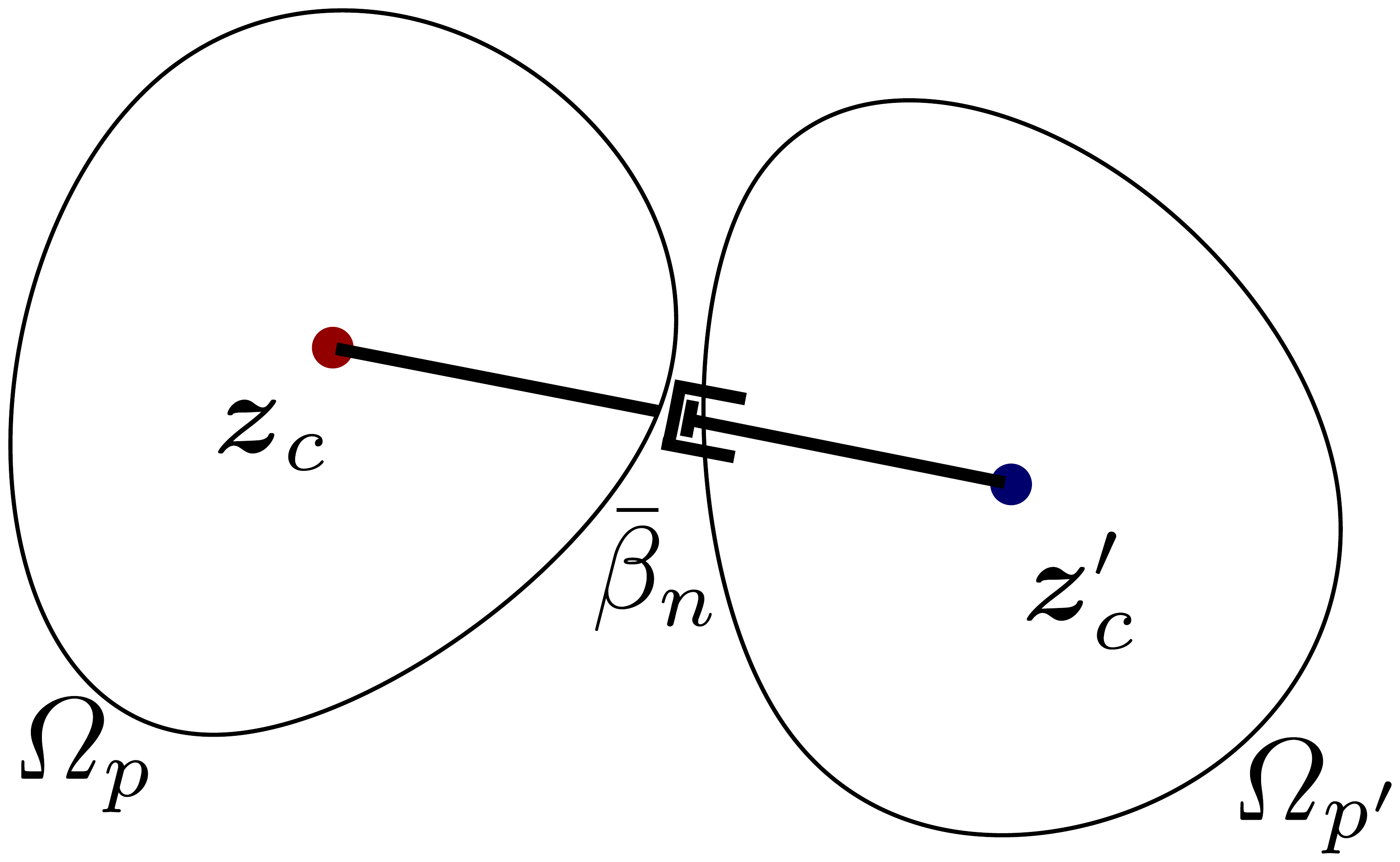}
	\caption{Model of damping acting between the particle centers.}\label{fig:discretecontact2}  
\end{figure}

Since $\bar{\bF}(\bx'_c, \bx_c)$ is the force density acting on the particle center, the force density on the individual nodes $\bx\in \varOmega_p$ is simply given by
\begin{align}\label{eq:dampingForce2}
	\bF_d(\bx) = \bar{\bF}_d(\bx'_c, \bx_c)\,.
\end{align}
This completes the description of the contact. Next, we briefly describe the numerical discretization and provide key information about the parameters for various simulations in \autoref{s:numeric}.

\paragraph{Remark on inter-particle contact}
Clear from the description of the model; the contact acts between the discretized nodes of two bodies. The criteria that two nodes of opposing bodies will have contact is based on the contact radius; only when the two points are within $R_c$ distance will they interact. Thus, the explicit description of the particle's boundary and, therefore, explicit formulation of contact based on the shape of the particle is not needed. The model naturally accounts for the shape effects on the contact interaction. It is possible that a node of a particle can have contact with more than one node of the opposite particle. This can happen for two reasons: 1) the shape of the particle itself (concave shape), and 2) the particle's deformation resulting in more than two nodes of a particle coming close to the node of the opposing particle. The model dynamically accounts for the shape change associated effects on the inter-particle interaction.

\section{Numerical discretization}\label{s:peridemNumericImp}
In this section, we provide the implementation details of the model discussed in previous section. We first discuss the meshless discretization of particles and write the discretized equation of motion. Next, we highlight how the contact parameters are calculated. For readers interested in the further details of the implementation, we refer to the open-sourced PeriDEM library\footnote{\url{https://github.com/prashjha/PeriDEM}}.

\subsection{Meshless discretization and discrete equation of motion}\label{ss:peridemImp}
\begin{figure}[!htb] 
	\centering
	\includegraphics[width=0.6\textwidth]{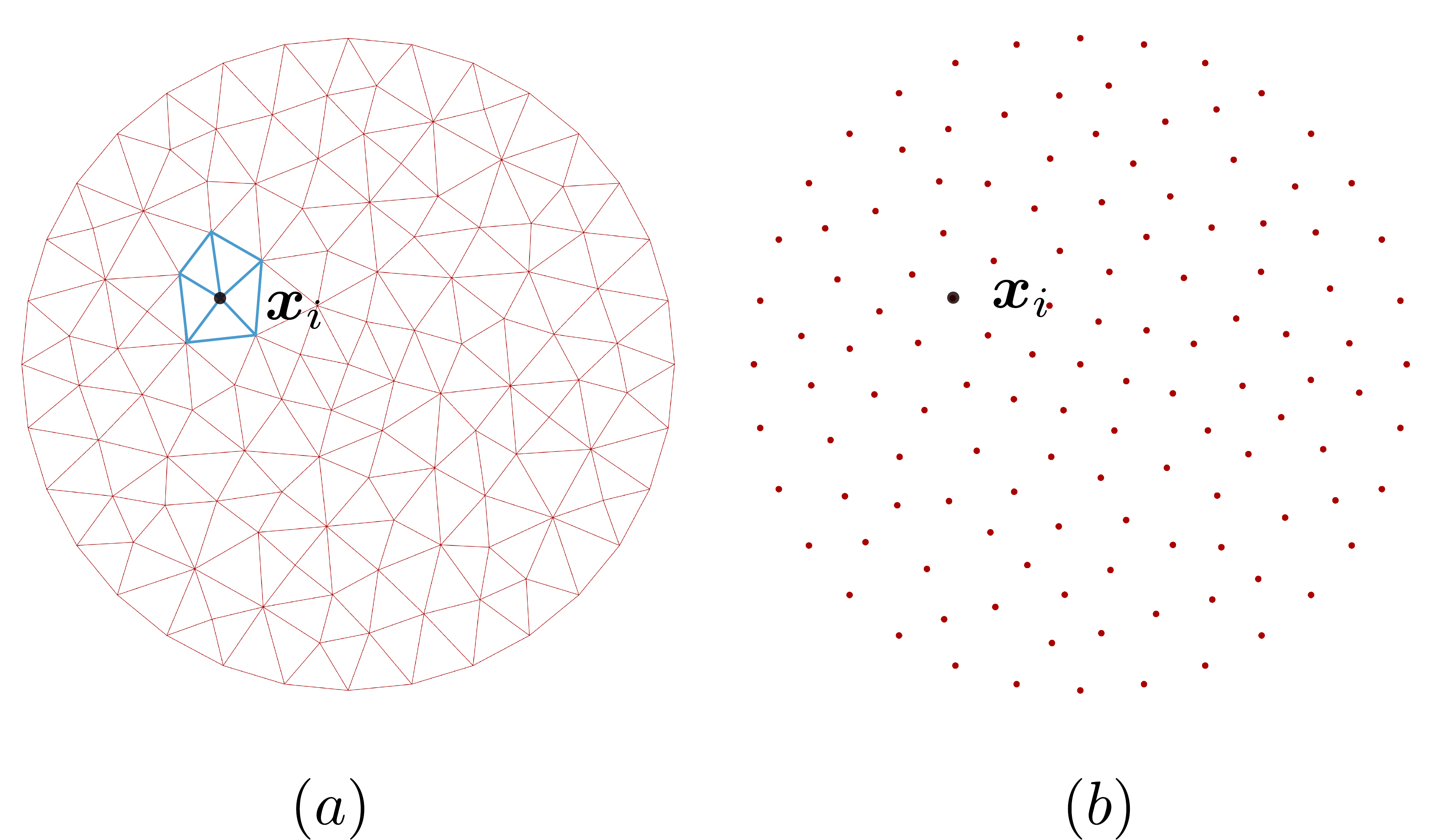}
	\caption{Discretization of the particle. {\bf (a)} The typical unstructured mesh of the particle. {\bf (b)} Corresponding meshless discretization. In {\bf (a)} and {\bf (b)}, we show the generic vertex $\bx_i$. In {\bf (a)}, the elements which have $\bx_i$ as the vertex are highlighted. }\label{fig:particleMesh}  
\end{figure}

Peridynamics is typically discretized using a meshless method (also referred to as meshfree); the meshfree discretization consists of the set of pairs of nodes $\bx_i$ and the nodal volume $V_i$. The sum of volumes $\sum V_i$ is equal to the volume of the domain. We utilize the Gmsh library \citet{geuzaine2009gmsh} for triangulation of the particle and wall. From the unstructured mesh, we obtain the meshless discretization, see \autoref{fig:particleMesh}. To be more precise: consider a $2$-dimensional problem with the mesh consisting of triangular elements. There are two approaches to obtain the meshless discretization: 1) taking the center of each triangle element as the node and the volume of element as the nodal volume, or 2) taking the vertices of the triangle as the node and computing the volume of each vertex from the interpolation function. We follow the second approach. Suppose $\phi_i$, $i \in \{1, 2, ..., N\}$, is the interpolation function associated with the vertex $i$. Also let $N_i$ is the list of elements $e$ in the mesh that has node $i$ as the vertex. Then the volume represented by the vertex $i$ is given by
\begin{align}\label{eq:nodalVol}
	V_{i} = \sum_{e \in N_i} \int_{T_e} \phi_i(\bx) \rmd\bx\,,
\end{align}
where $T_e$ is the element domain. 

For the temporal discretization of \autoref{eq:pdmotion}, we consider a central-difference scheme. This results in the following equation governing the evolution of the displacement of the node $i$:
\begin{align}\label{eq:pdmotionDisc}
\rho_i \frac{\bu_i^{n+1} - 2\bu_{i}^n + \bu_i^{n-1}}{\Delta t^2}  = \bF_i + \bF_{i,ext}\,,
\end{align}
where $\rho_i, \bu_i, \bF_i, \bF_{i, ext}$ denotes density, displacement, internal force density, and contact and external force densities at the node $i$. $\bF_{i, ext}$ includes the contact forces due to the contact of the particle associated with the node $i$ with the neighboring particles. $\bu_i^n$ denotes the displacement of node $i$ at time $t_n = n\Delta t$. $\bF_i$ is the approximation of the peridynamics force density on node $i$ and is given by
\begin{align}\label{eq:pdForceApprox}
  \bF_i = \sum_{\substack{j, |\bx_j - \bx_i| < \epsilon, \\
    \bx_j \neq \bx_i}} (\bT_{\bx_i}(\bx_j) - \bT_{\bx_j}(\bx_i)) V_j\,,
\end{align}
where $T_{\bx_j}(\bx_i)$ is the approximation of $T_{\by}(\bx)$, see \autoref{eq:state_Tx3}. It depends on $\theta_{\bx_i}$ and $m_{\bx_i}$ which are approximately computed as follows:
\begin{align}\label{eq:mxDisc}
  m_{\bx_i} &= \sum_{\substack{j, |\bx_j - \bx_i| < \epsilon, \\
      \bx_j \neq \bx_i}} |\bx_j - \bx_i|^2 J(|\bx_j - \bx_i|/\epsilon) V_j\,, \notag \\
  \theta_{\bx_i} &= \frac{3}{m_{\bx_i}}\sum_{\substack{j, |\bx_j - \bx_i| < \epsilon, \\
      \bx_j \neq \bx_i}} h(s(\bx_j, \bx_i),t) |\bx_j - \bx_i| (|\bz(\bx_j) - \bz(\bx_i)| - |\bx_j - \bx_i|) J(|\bx_j - \bx_i|/\epsilon) V_j\,.
\end{align}
This completes the description of the discretization of the equation of motion. The following section provides details about the contact force calculation.

\subsection{Contact parameters}
We define the mesh size $h$ as follows:
\begin{align}\label{eq:meshSize}
	h = \min_{\substack{\bx_i, \bx_j \\
			\bx_i \neq \bx_j }} |\bx_i - \bx_j|\,.
\end{align}
In all simulations we have fixed the contact radius using $R_c = 0.95 h$, where $h$ is specific to the different simulation. For the spring modulus $K_n$, we use the formula \citet{behzadinasab2018peridynamics, silling2005meshfree}
\begin{align}\label{eq:springModulus}
	K_n = \frac{18 \kappa}{\pi \epsilon^5}\,,
\end{align}
where $\epsilon$ is the horizon and $\kappa$ is the bulk modulus. In the case when the contacting bodies have different bulk moduli, $\kappa_1$ and $\kappa_2$, we define $K_n$ using an effective bulk modulus $\kappa_{eff}$; see \autoref{eq:effk} for $\kappa_{eff}$ formula.

Now it remains to specify the damping parameter. In this work, we apply the damping between the  particle centers described in \autoref{ss:damping2}. In all simulations, we have fixed $\bar{C} = 100$. The value of the parameter $\bar{\varepsilon}_n$ is specific to the numerical example and will be discussed when describing the setup.

\paragraph{Particle-wall damping force}
To apply the damping between particle $\varOmega_p$ and wall $\varOmega_w$, we use the damping mechanism described in \autoref{ss:damping2} while treating the individual nodes $\bx_i\in \varOmega_w$ of wall as the opposite particle $\varOmega_{p'}$ in the description in \autoref{ss:damping2}.

\paragraph{Prevention of self-penetration} 
To prevent the self-penetration in the solid $\varOmega$, we apply the normal contact force between nodes $\bx_i$ and $\bx_j$ when the peridynamics bond between them is broken. The contact force formula is the same as in \autoref{eq:normalForce} where $K_n$ is computed from the bulk modulus as in \autoref{eq:springModulus}. The contact radius, as in the case of inter-particle contact, is fixed by $R_c = 0.95 h$.

\section{Numerical tests}\label{s:numeric}
In this section, we apply the model to examine two and multi-particle systems thoroughly and highlight key features. First, we apply the model to a two-particle setup and show the damping effects. Next, we study the effect of mesh size on model behavior. The model is not restricted to only circular particles; this is shown through examples with hexagon-shaped and non-convex particles. As the concluding example, we use the model to simulate a compressive test of particulate media comprising 500+ hexagon-shaped and circular particles.

\subsection{Two-particle test}\label{ss:twop}
\begin{figure}[!htb] 
  \centering
  \includegraphics[width=0.4\textwidth]{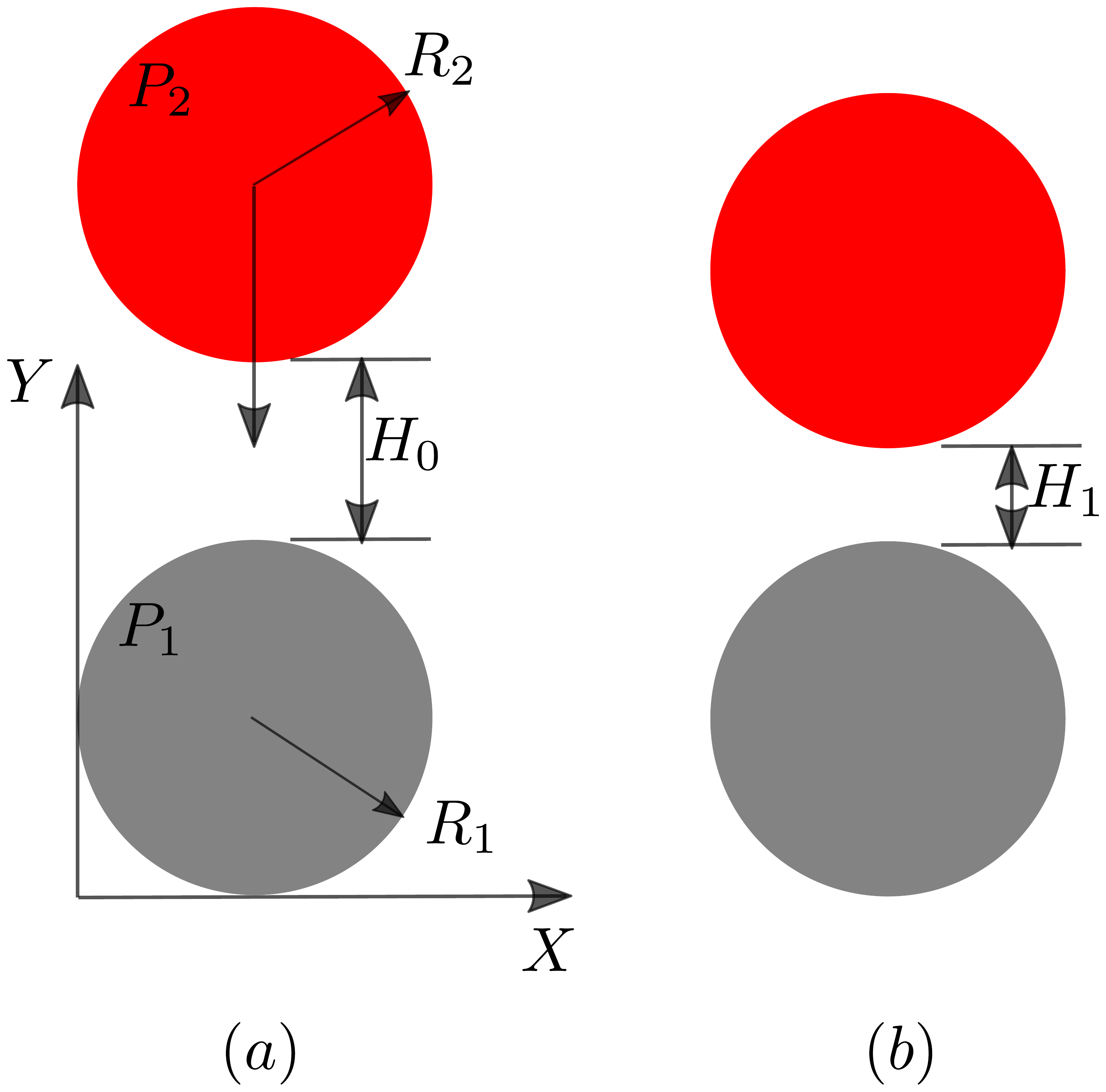}
  \caption{Setup for the two-particle test. {\bf (a)} Initial state. The particle at the bottom is fixed and rigid, whereas the particle on top is dropped from the height $H_0$ with zero initial velocity. {\bf (b)} Later time after the first contact when the top particle has bounced to the maximum height. $H_1$ is the relative distance in this state.}\label{fig:twoPSetup}  
\end{figure}

As the simplest example, we consider two particles in which the particle at the bottom is fixed and rigid, and the particle on the top is falling due to gravity; see the setup in \autoref{fig:twoPSetup}. We study the effect of the damping parameter $\bar{\varepsilon}_n$ on the rebound height after the first contact. We fix the gravitational acceleration $g = 10$ m/s$^2$ in the downward direction. If $H_0$ is the initial distance between the two particles and $H_1$ is the maximum distance after the first contact, the coefficient of restitution (CoR) is given by
\begin{align}\label{eq:cor}
  C_R = \sqrt{\frac{H_1}{H_0}}\,.
\end{align}
$C_R = 1$ implies perfectly elastic collision, whereas $C_R < 1$ implies the system's loss of energy due to the damping. $C_R$ is specific to the two materials in contact, \ie, it depends on the material properties of bodies in touch. The $C_R$ table for different pairs of materials is used to calibrate the damping coefficient $\bar{\varepsilon}_n$, see \citet{desai2017tribosurface, desai2019rheometry}.

To demonstrate the effect of damping on the coefficient of restitution, we consider two different materials and perform tests where we combine the material type and vary the radius of the top and bottom particles. In \autoref{tab:materials}, we list the properties of two materials. Particles are discretized with the mesh size $h = 0.1423$ mm. The horizon for the nonlocal model is fixed to $\epsilon = 0.6$ mm. The total time of the simulation is $T = 0.04$ s. The timestep size largely depends on the spring constant $K_n$. We consider $\Delta t = 0.2, 0.02$ $\mu$s for tests with material M1 and M2 respectively. For the material pair (M1, M2), we consider $\Delta t = 0.1$ $\mu$s. 

\begin{figure}[!htb] 
  \centering
  \includegraphics[width=0.9\textwidth]{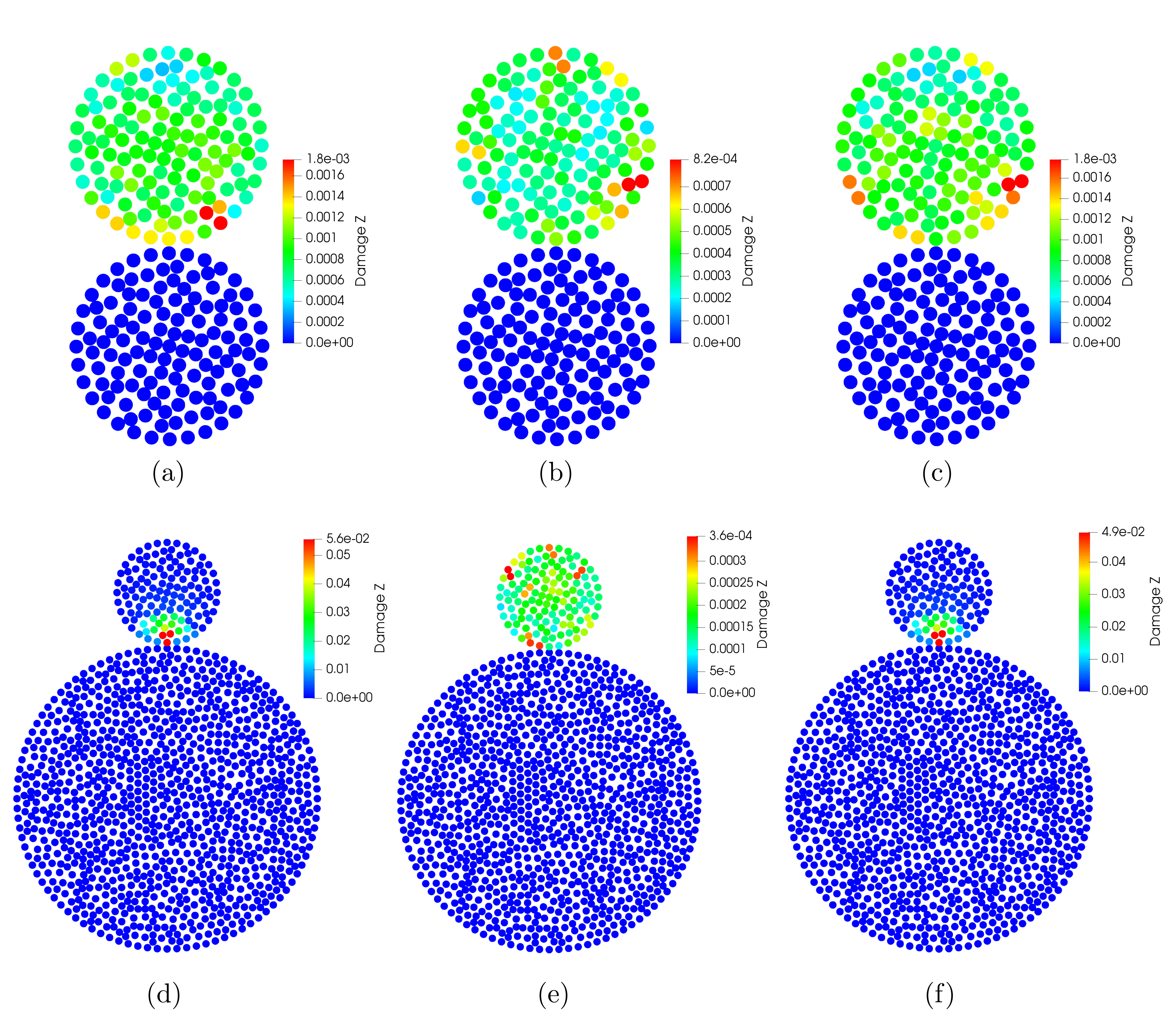}
  \caption{Plot of damage within particles near the contact time $t = 0.0044$ s. Damage function $Z$ is defined in \autoref{eq:damage}. $Z(\bx)<1$ implies no fracture (elastic deformation) whereas $Z(\bx)\geq 1$ implies one or more bonds in the neighborhood of point $\bx$ is broken. {\bf (a)} and {\bf (b)} correspond to the tests 1 and 2 with $\bar{\varepsilon}_n = 0.95$ in \autoref{tab:corResults}. {\bf (c)}, {\bf (d)}, {\bf (e)}, {\bf (f)} corresponds to the tests 9, 6, 8, 10 respectively in \autoref{tab:corResults2}.}\label{fig:twoPZPlot1}  
\end{figure}

In \autoref{tab:corResults}, we list the values of $C_R$ for different damping parameter $\bar{\varepsilon}_n$ for particles of material M1 and radius $R_1 = R_2 = 1$ mm. Naturally in the absence of damping we have elastic contact in which the top particle rebounds to the same height resulting in $C_R = 1$. In \autoref{tab:corResults2}, we list the values of $C_R$ for various tests for the case of elastic contact and contact with $\bar{\varepsilon}_n = 0.95$. For the fixed parameter $\bar{\varepsilon}_n$, the effect of damping is smaller in the material with smaller strength. In all cases decrease in $\bar{\varepsilon}_n$ implies increase in the damping and therefore decrease in $C_R$. We present the simulation results for the subset of the cases in \autoref{tab:corResults} and \autoref{tab:corResults2} in \autoref{fig:twoPZPlot1}.

\begin{table}[!htb] 
  \centering
  \begin{tabular}{|c|c|c|c|c|}
    \hline
    Material set & $\rho$ & $K$ & $G$ & $G_c$ \\
    \hline
    \hline
    M1 & 1200 Kg/m$^3$ & 0.0216 GPa & 0.01296 GPa & 50 J/m$^2$ \\
    M2 (PMMA Glass) & 1200 Kg/m$^3$ & 2 GPa & 1.2 GPa & 500 J/m$^2$ \\
    \hline
  \end{tabular}
  \caption{Two sets of materials. Here $\rho$ denotes density, $K$ bulk modulus, $G$ shear modulus, and $G_c$ critical energy release rate.}\label{tab:materials}
\end{table}

\begin{table}[!htb] 
  \centering
  \begin{tabular}{|c|c|c|}
    \hline
    Test & $\bar{\varepsilon}_n$ & $C_R$ \\
    \hline
    \hline
    1 & 1 & 1 \\
    2 & 0.95 & 0.946 \\
    3 & 0.9 & 0.893 \\
    4 & 0.85 & 0.845 \\
    5 & 0.8 & 0.796 \\
    \hline
  \end{tabular}
  \caption{Coefficient of restitution for the two-particle test. Here, $R_1 = R_2 = 1$ mm and the material properties of both particles are same as M1 in \autoref{tab:materials}. The initial separation between the particles is $H_0 = 1$ mm. As $\bar{\varepsilon}_n$ decreases, $C_R$ decreases. 
  }\label{tab:corResults}
\end{table}

\begin{table}[!htb] 
  \centering
  \begin{tabular}{|c|c|c|c|c|}
    \hline
    Test & $(R_1, R_2)$ & Material pair & $C_R$ ($\bar{\varepsilon}_n = 1$) & $C_R$ ($\bar{\varepsilon}_n = 0.95$) \\
    \hline
    \hline
    6 & (3,1) & (M1, M1) & 1 & 0.935 \\
    7 & (1,1) & (M2, M2) & 1 & 0.744 \\
    8 & (3,1) & (M2, M2) & 1 & 0.716 \\
    9 & (1,1) & (M2, M1) & 1 & 0.925 \\
    10 & (3,1) & (M2, M1) & 1 & 0.914 \\
    \hline
  \end{tabular}
  \caption{Coefficient of restitution corresponding to the mixed cases, \ie, either with the different radii or different material properties. We consider elastic collision and collision with $\bar{\varepsilon}_n = 0.95$. In all tests we have $H_0 = 1$ mm. We see that for the fixed $\bar{\varepsilon}_n = 0.95$, the damping effect is stronger in the material with more strength (compare tests 6 and 8, 7 and 9, 6 and 10). }\label{tab:corResults2}
\end{table}

\begin{figure}[!htb] 
  \centering
  \includegraphics[width=0.9\textwidth]{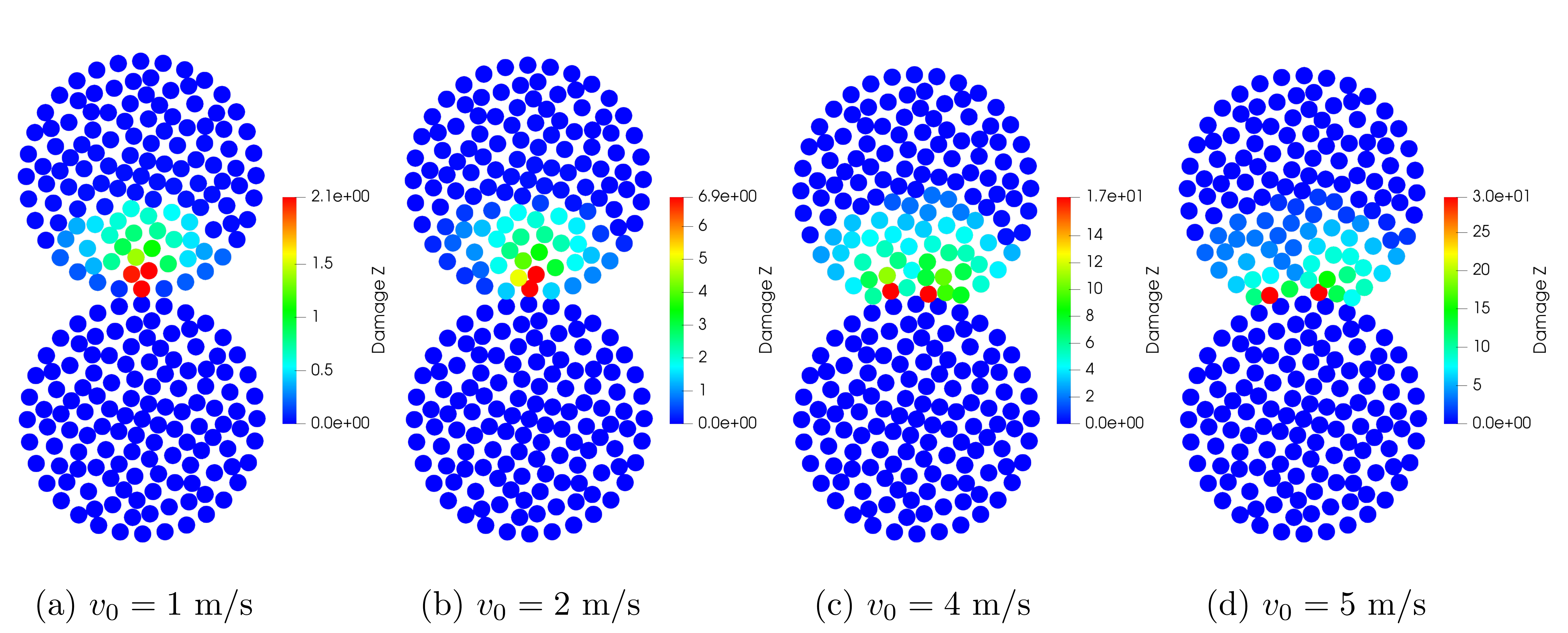}
  \caption{Plot of damage for two-particle test 2 in \autoref{tab:corResults}. $v_0$ is the initial velocity of the top particle in the downward direction. Final time and the time step for all four cases are $T = 0.001$ s and $\Delta t = 0.2\,\mu$s. For all cases, the fracture zone $FZ$ is present, \ie, there are nodes such that $Z(\bx) \geq 1$. }\label{fig:twoPFracture}  
\end{figure}

\begin{figure}[!htb] 
	\centering
	\includegraphics[width=0.9\textwidth]{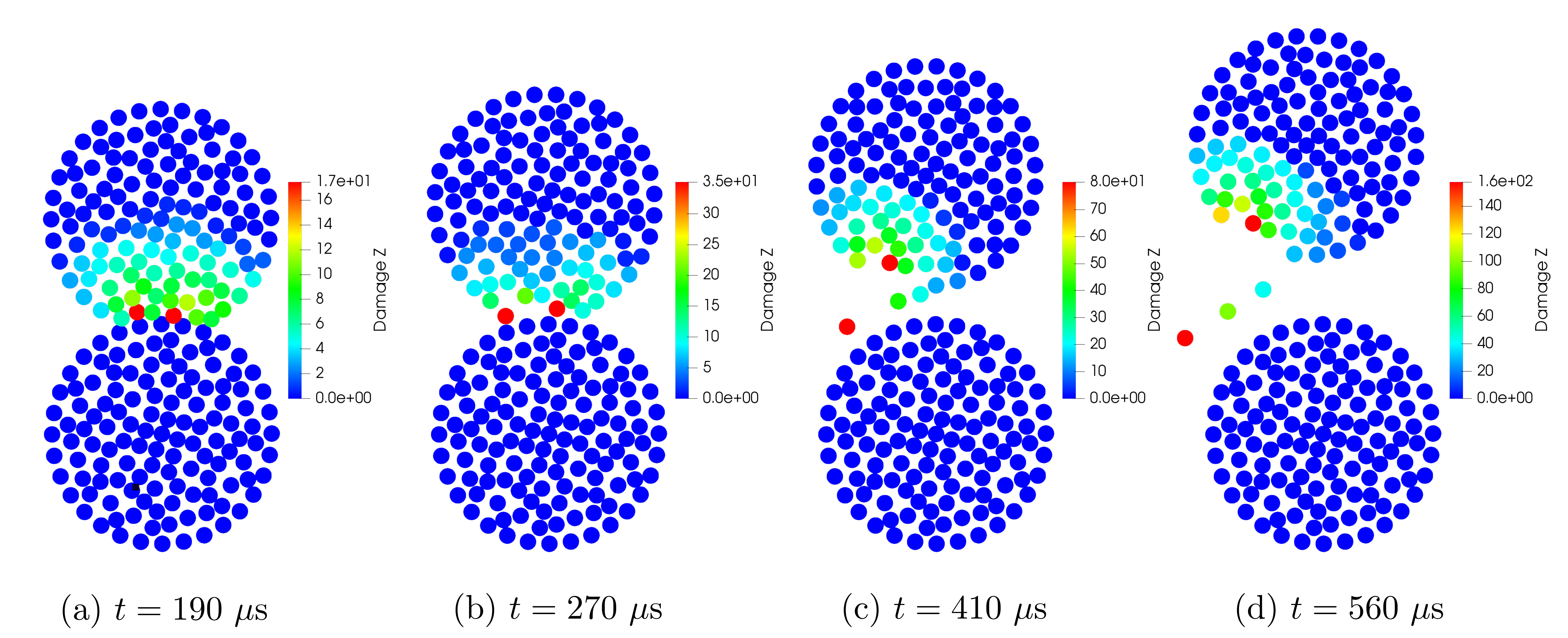}
	\caption{Particle evolution at 4 different times. The results correspond to the test 2 in \autoref{tab:corResults} with $v_0 = 5$ m/s.}\label{fig:twoPFractureEvol}  
\end{figure}

\begin{figure}[!htb] 
  \centering
  \includegraphics[width=0.9\textwidth]{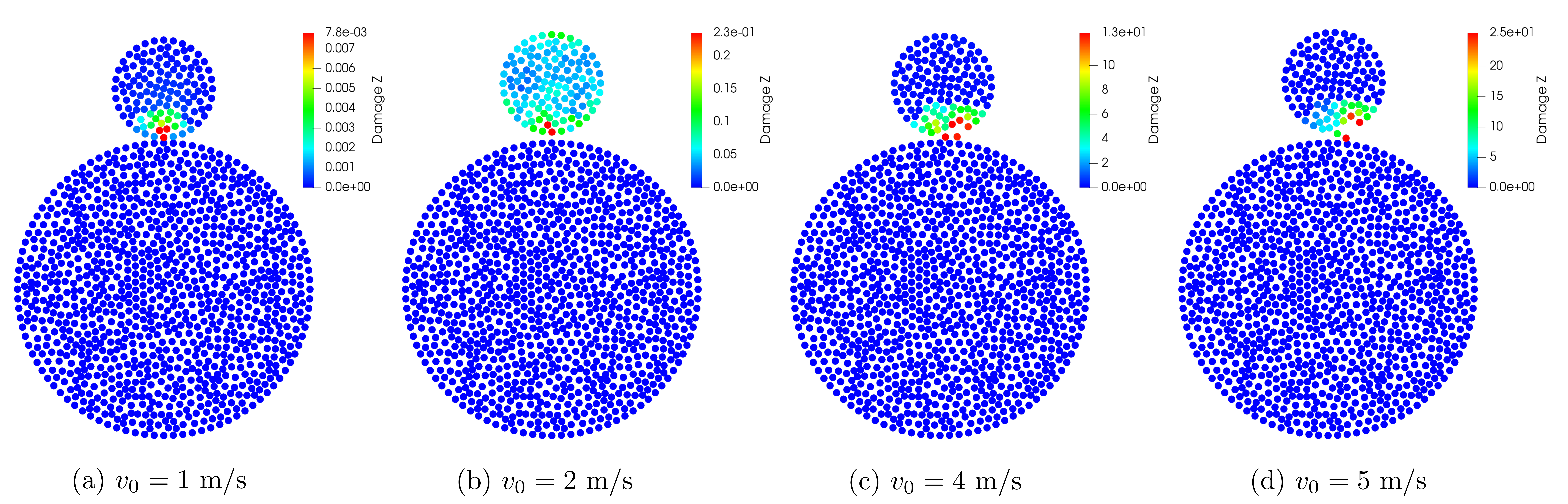}
  \caption{Plot of damage for two-particle test 10 ($\bar{\varepsilon}_n = 0.95$) in \autoref{tab:corResults2}. Final time and the time step for all four cases are $T = 0.001$ s and $\Delta t = 0.1\,\mu$s.}\label{fig:twoPFractureM12}  
\end{figure}

\begin{figure}[!htb] 
  \centering
  \includegraphics[width=0.9\textwidth]{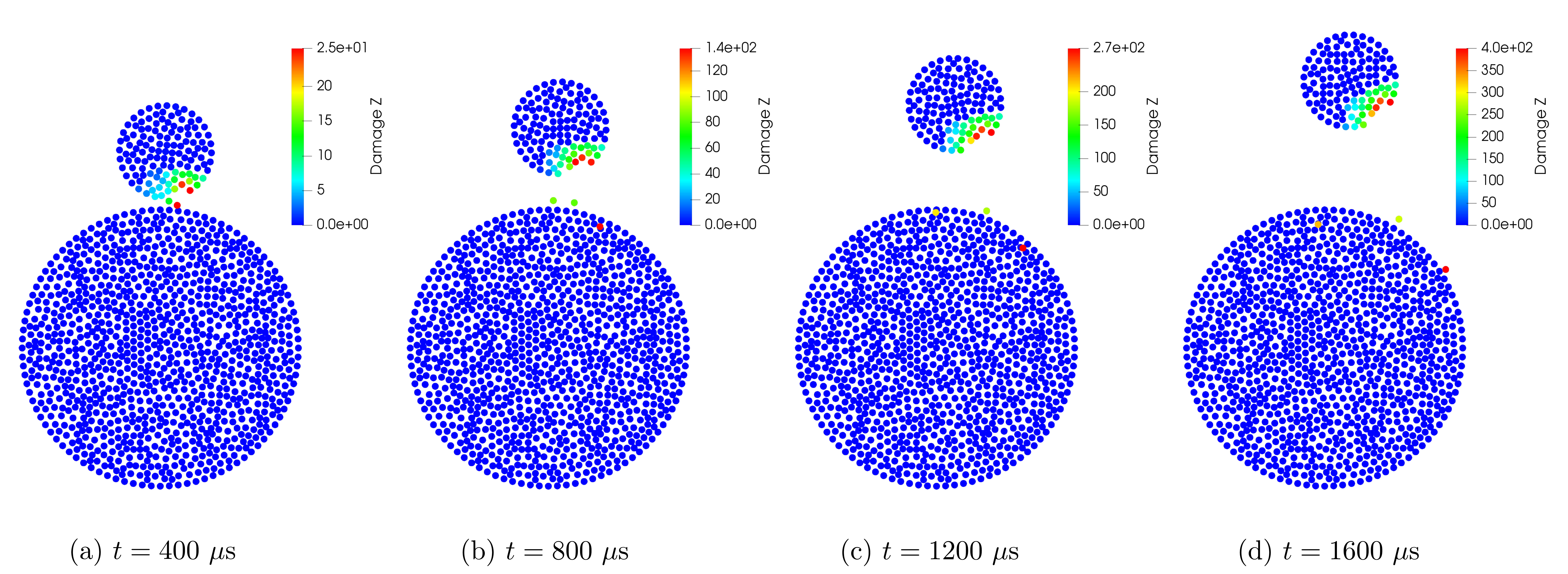}
  \caption{Particle evolution at 4 different times. The results correspond to the test 10 ($\bar{\varepsilon}_n = 0.95$) in \autoref{tab:corResults2} with $v_0 = 5$ m/s.}\label{fig:twoPFractureEvolM12}  
\end{figure}

\paragraph{Fracture simulations}
We assign initial downward velocity $v_0$ to the top particle in \autoref{fig:twoPSetup}. As we increase the initial velocity $v_0$, the damage on particles where they contact should increase leading to failure at high enough velocities. In \autoref{fig:twoPFracture}, we show the plot of damage just after the contact for the different values of $v_0$. Here the tests are similar to the test 2 in \autoref{tab:corResults} with the only difference in the current tests is that the top particle is assigned nonzero initial velocity. The evolution of the top particle for the case of $v_0 = 5$ m/s at four times is presented in \autoref{fig:twoPFractureEvol}. For the same range of velocities, we considered test 10 in \autoref{tab:corResults2} with $\bar{\varepsilon}_n = 0.95$. The damage plots for this case are shown in \autoref{fig:twoPFractureM12} and evolution of the top particle for the case of $v_0 = 5$ m/s is presented in \autoref{fig:twoPFractureEvolM12}. 

\subsubsection{Mesh effect}\label{ss:meshEffects}
\begin{figure}[!htb] 
	\centering
	\includegraphics[width=0.35\textwidth]{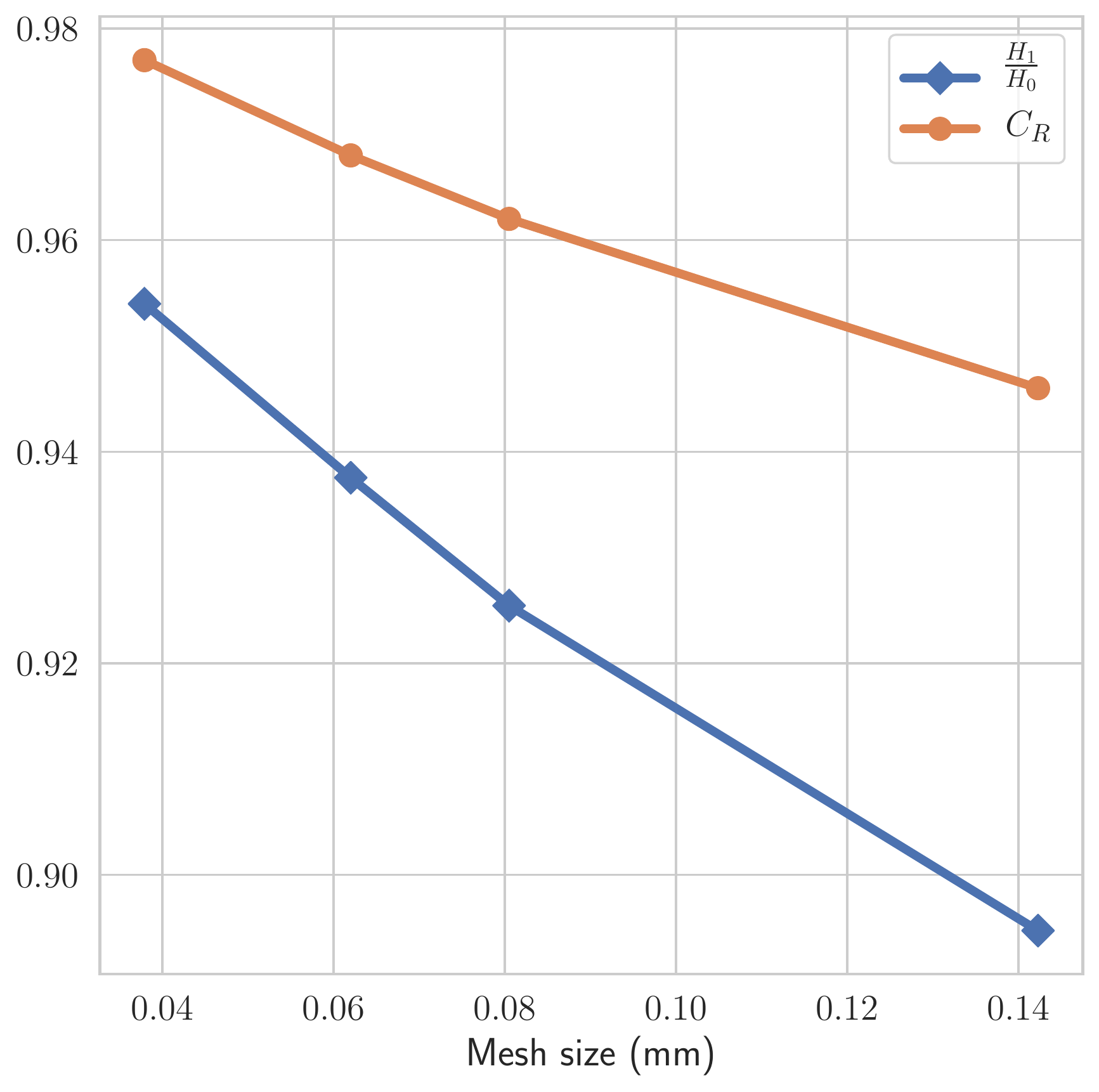}
	\caption{$C_R$ and $H_1/H_0$ for different mesh sizes. The $C_R$ is increasing with the decreasing mesh size.}\label{fig:twoPMeshPlot3}  
\end{figure}

\begin{figure}[!htb] 
  \centering
  \includegraphics[width=0.9\textwidth]{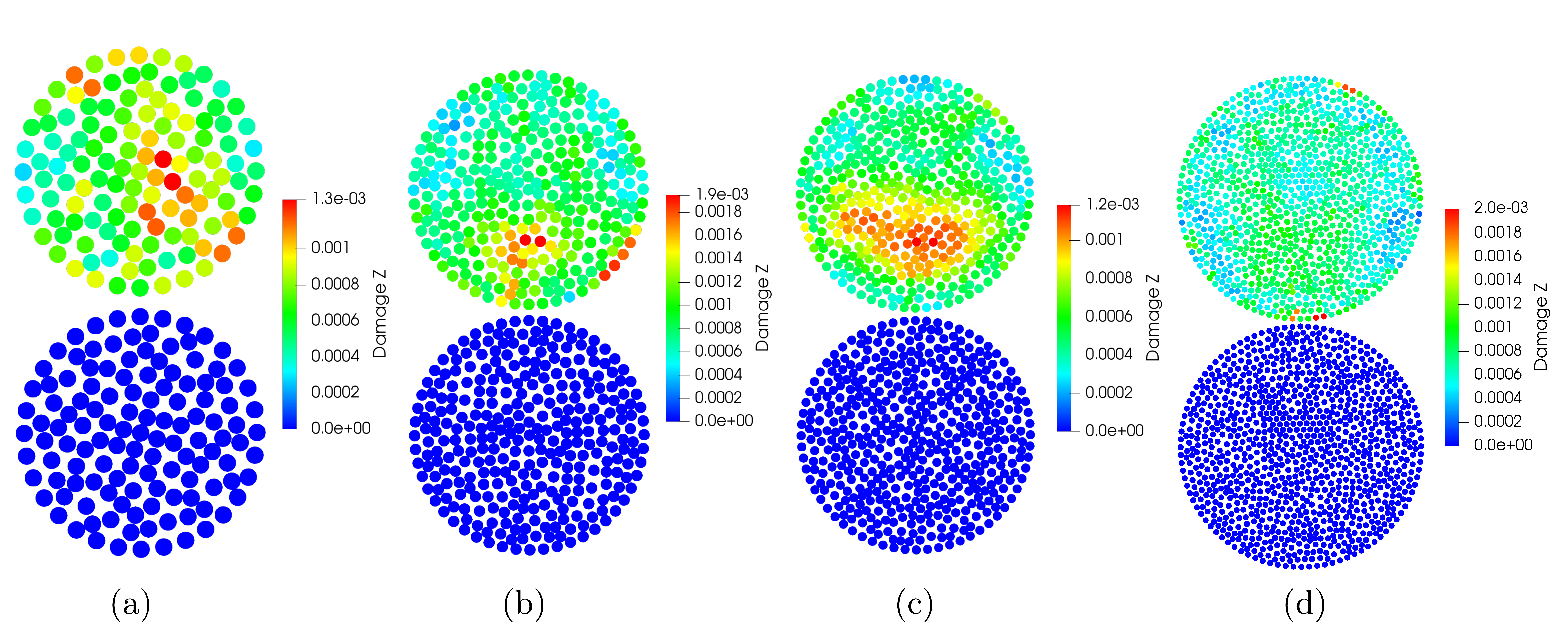}
  \caption{Plot of damage for the tests in \autoref{tab:corMesh}. Note that the magnitude of damage is of the same order for all the four simulations.
   }\label{fig:twoPZPlot3}  
\end{figure}

Consider test 2 in \autoref{tab:corResults}. To see how the model behaves with different mesh sizes, we consider four meshes of decreasing mesh size and horizon and record the $C_R$ while keeping the fixed damping coefficient $\bar{\varepsilon}_n = 0.95$. We list the $C_R$ values in \autoref{tab:corMesh}. In \autoref{fig:twoPMeshPlot3}, we plot $C_R$ and $H_1/H_0$ for four mesh sizes. For all four cases, we show the damage near the contact time in \autoref{fig:twoPZPlot3}. The contact radius depends on the mesh size via $R_c = 0.95 h$. Thus with the decreasing mesh size, the size of the contact neighborhood decreases. The duration for which the damping is effective depends on the contact neighborhood. Therefore with the smaller mesh size, the reduced damping effect (implying increased $C_R$) is seen in \autoref{tab:corMesh} and \autoref{fig:twoPMeshPlot3}. Another contributing factor to the trend in \autoref{tab:corMesh} could be the peridynamics; we take the horizon proportional to the mesh size, and therefore as the mesh size changes, the horizon changes. Due to the nature of the method (the contact is defined on the discretization and $R_c = 0.95h$), it is expected that the mesh size will influence the contact dynamics. Further studies could help identify the major factors and possibly modify the contact parameters such that the mesh effect is minimal. It may also be possible that the limit of $C_R$ as $h \to 0$ in \autoref{fig:twoPMeshPlot3} is not $1$ but some fixed smaller number.

\begin{table}[!htb] 
  \centering
  \begin{tabular}{|c|c|c|c|c|c|c|}
    \hline
    Test & Mesh size (mm) & Horizon (mm) & $\Delta t$ ($\mu$s) & $H_0$ (mm) & $C_R$ ($\bar{\varepsilon}_n = 1$) & $C_R$ ($\bar{\varepsilon}_n = 0.95$) \\
    \hline
    \hline
    1 & 0.1423 & 0.6 & 0.2 & 1 & 1 & 0.946 \\
    2 & 0.0805 & 0.375 & 0.1 & 1 & 1 & 0.962 \\
    3 & 0.062 & 0.3 & 0.1 & 1 & 1 & 0.968 \\
    4 & 0.0379 & 0.2 & 0.05 & 1 & 1 & 0.977 \\
    \hline
  \end{tabular}
  \caption{$C_R$ for the case when particles have the same radius $R = 1$ mm and have same material properties M1. We note that as the mesh size decreases, $C_R$ increases.}\label{tab:corMesh}
\end{table}

\subsection{Two-particle with wall} \label{ss:twoPWall}
\begin{figure}[!htb] 
  \centering
  \includegraphics[width=0.35\textwidth]{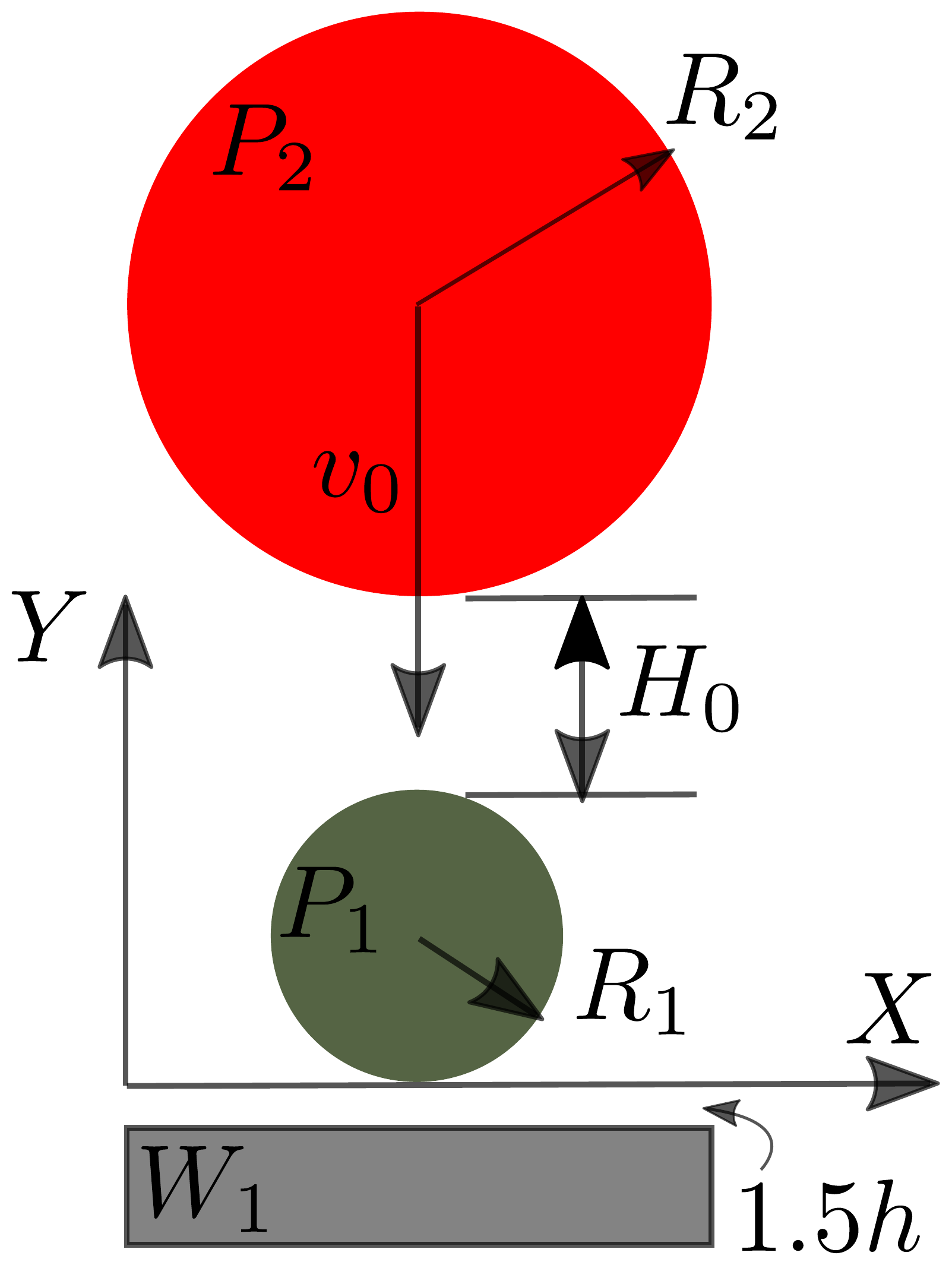}
  \caption{Schematics of the two-particle with wall test. Here both particles are falling freely due to the downward gravity $g=10$ m/s$^2$. The wall is fixed in place and is assumed to be rigid solid. $P_2$ is given an initial velocity of $v_0$ downwards.}\label{fig:twoPWSetup}  
\end{figure}

We consider a setup similar to \autoref{ss:twop} but now with a fixed wall below the bottom particle.  The configuration shown in \autoref{fig:twoPWSetup} consists of two particles of different radii and materials. The bottom particle is free-falling, whereas the top particle is assigned an initial downward velocity $v_0$.

\begin{figure}[!htb]   
  \centering
  \includegraphics[width=0.9\textwidth]{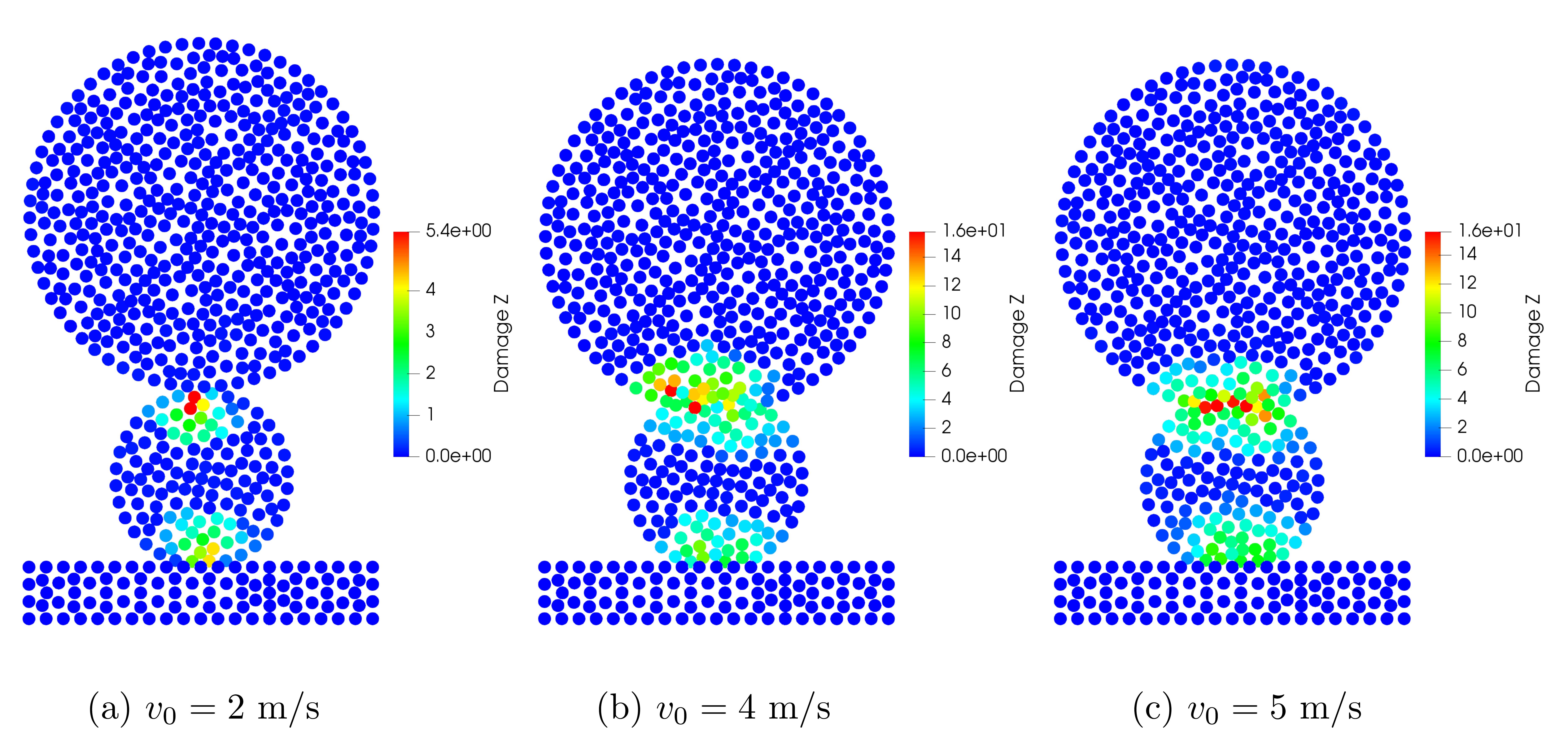}
  \caption{The plot of damage when particles and wall have the same material properties. Since both particles have the same strength and critical energy release rate, both break.}\label{fig:twoPWFracture}  
\end{figure}

\begin{figure}[!htb] 
  \centering
  \includegraphics[width=0.9\textwidth]{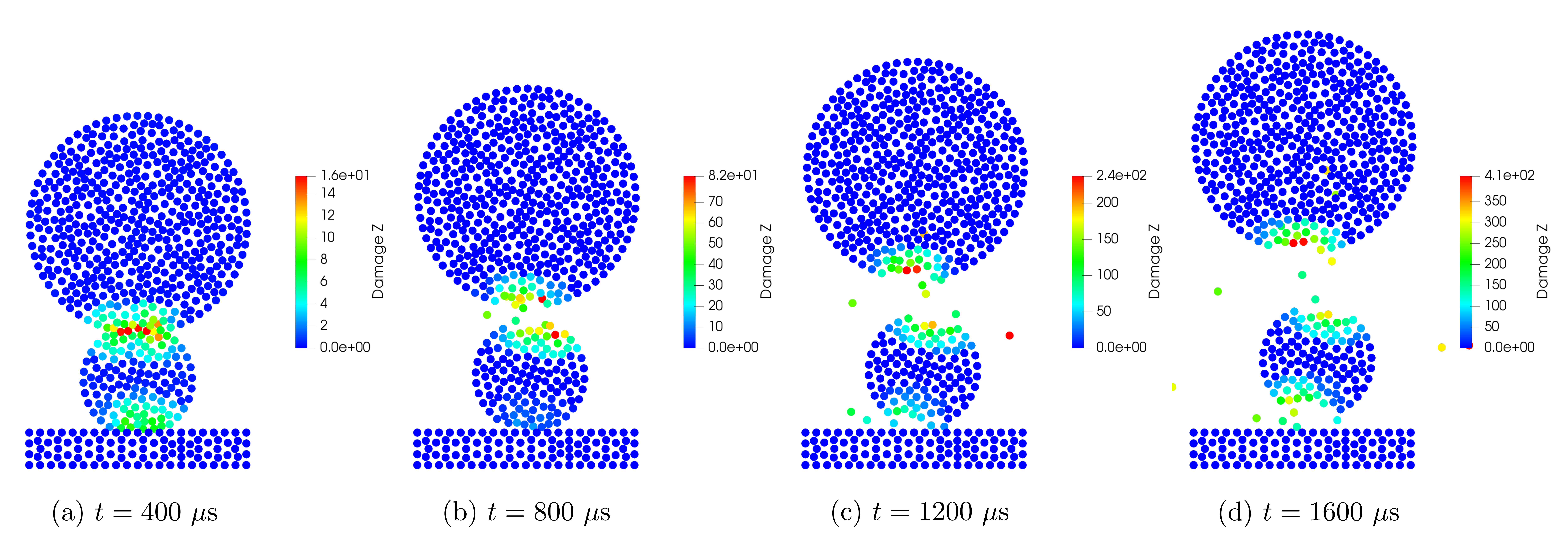}
  \caption{Evolution of the system for the test with same material properties and $v_0 = 5$ m/s.}\label{fig:twoPWFractureEvol}  
\end{figure}

\paragraph{Both particles and wall made-up of the same material M1} We take $R_1 = 1$ mm, $R_2 = 2$ mm, $H_0 = 1$ mm. The mesh size is $h = 0.1423$ mm and horizon is $\epsilon = 0.6$ mm. Final simulation time is $T = 0.04$ s and the time step is $\Delta t = 0.1$ $\mu$s. Damping parameter $\bar{\varepsilon}_n$ is fixed to $0.95$. We consider three different initial velocities $v_0 = 2, 4, 5$ m/s. The damage at the impact time for all three cases are shown in \autoref{fig:twoPWFracture}. Evolution of the system for the case when $v_0 = 5$ m/s is shown in \autoref{fig:twoPWFractureEvol}. Since both particles have the same material strength, the fracture is seen on both. The region containing points with broken bonds in the neighborhood (fracture zone $FZ$, see \autoref{eq:damageZone}) increases with the increasing $v_0$. 

\begin{figure}[!htb] 
  \centering
  \includegraphics[width=0.9\textwidth]{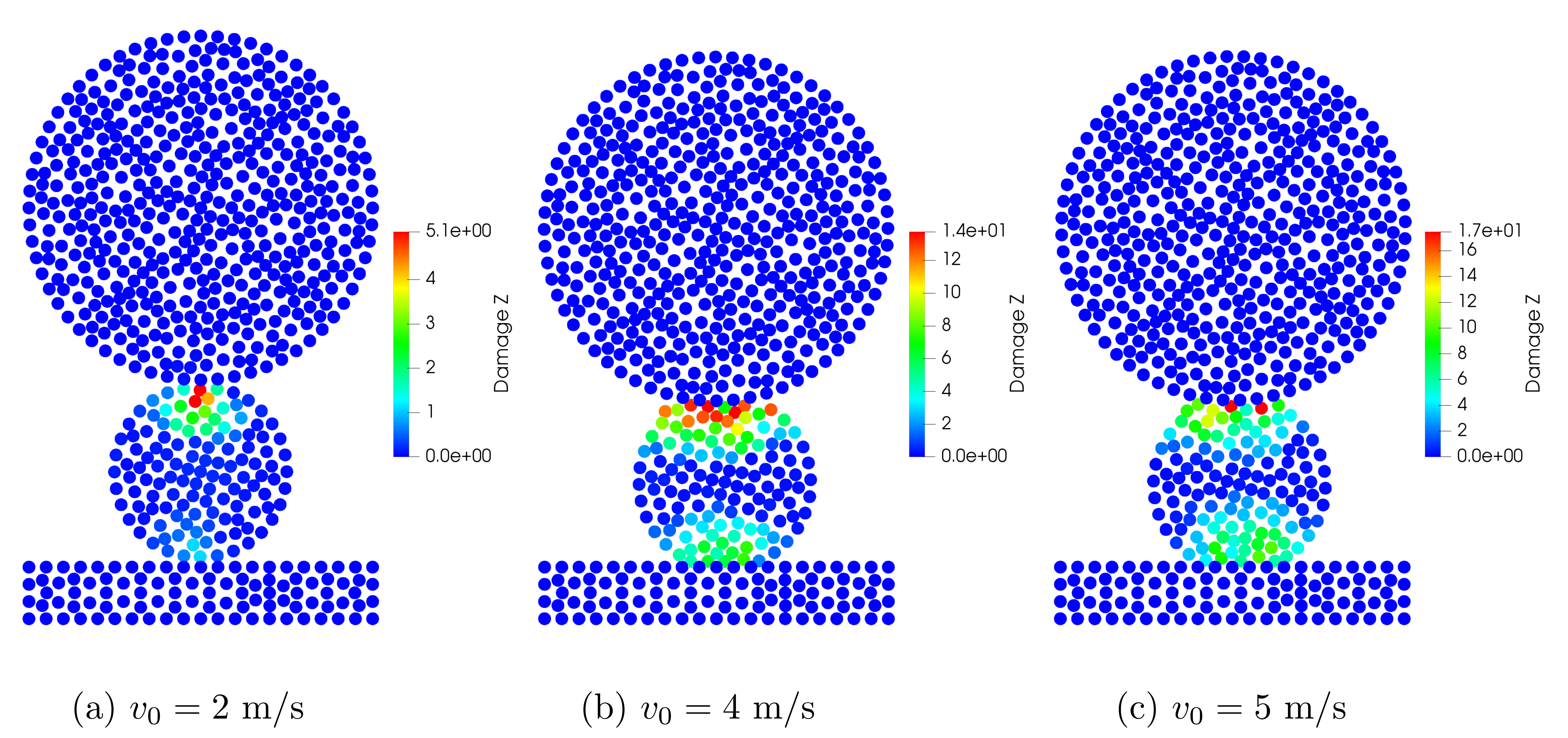}
  \caption{The plot of damage when the particles and the wall have different material properties. The wall and top particle are of material M2, whereas the bottom particle is of M1. Since the top particle has higher strength and higher critical energy release rate than the bottom particle, only the bottom particle breaks.}\label{fig:twoPWFractureM12}  
\end{figure}

\begin{figure}[!htb] 
  \centering
  \includegraphics[width=\textwidth]{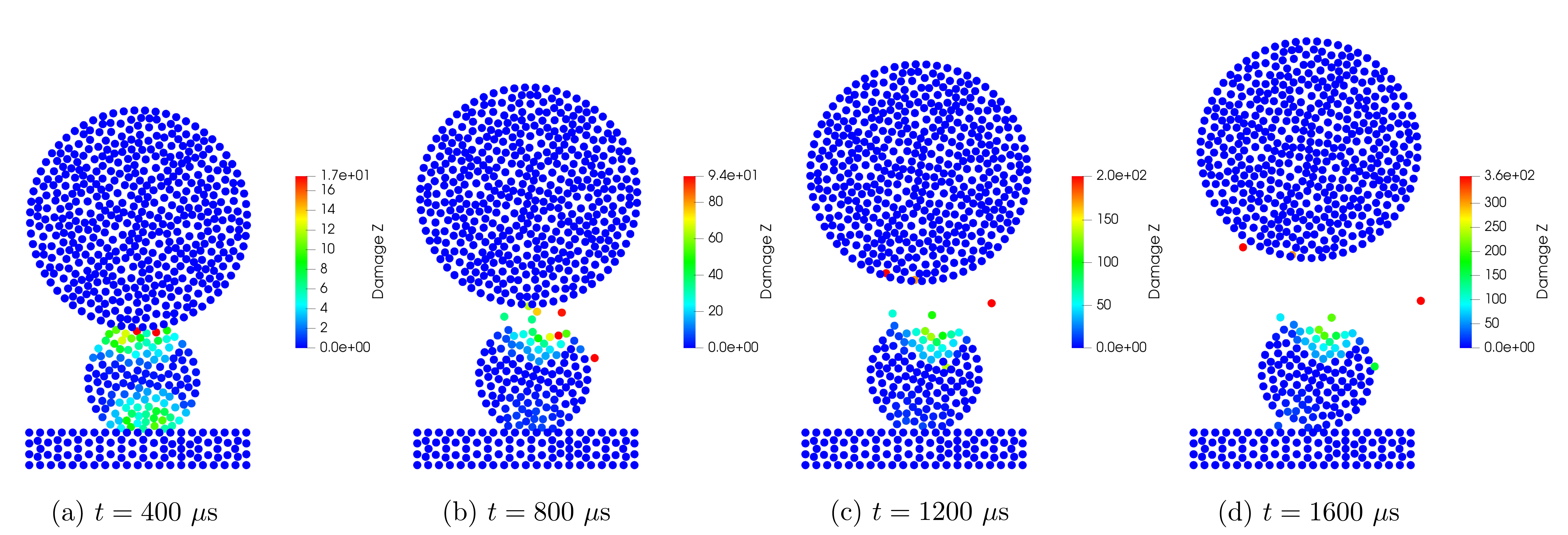}
  \caption{Evolution of the system for the test with different material properties and initial velocity $v_0 = 5$ m/s. Note that the top particle is undamaged (no nodes with $Z\geq1$) as it has higher strength. }\label{fig:twoPWFractureEvolM12}  
\end{figure}

\paragraph{Particles and wall made-up of different materials} For the top particle and the fixed wall, we consider material M2 with higher strength. For the particle in between the wall and top particle, we consider material M1. Other parameters are the same as before. The damage at the impact time for three different initial velocities $v_0 = 2,4,5$ m/s are shown in \autoref{fig:twoPWFractureM12}. Evolution of the system for the case when $v_0 = 5$ m/s is shown in \autoref{fig:twoPWFractureEvolM12}. Note that only the bottom particle sustains damage while the top particle remains intact. This is expected as the top particle has higher strength.

\subsection{Non-circular particles} \label{ss:noncircular}
\begin{figure}[!htb] 
  \centering
  \includegraphics[width=0.4\textwidth]{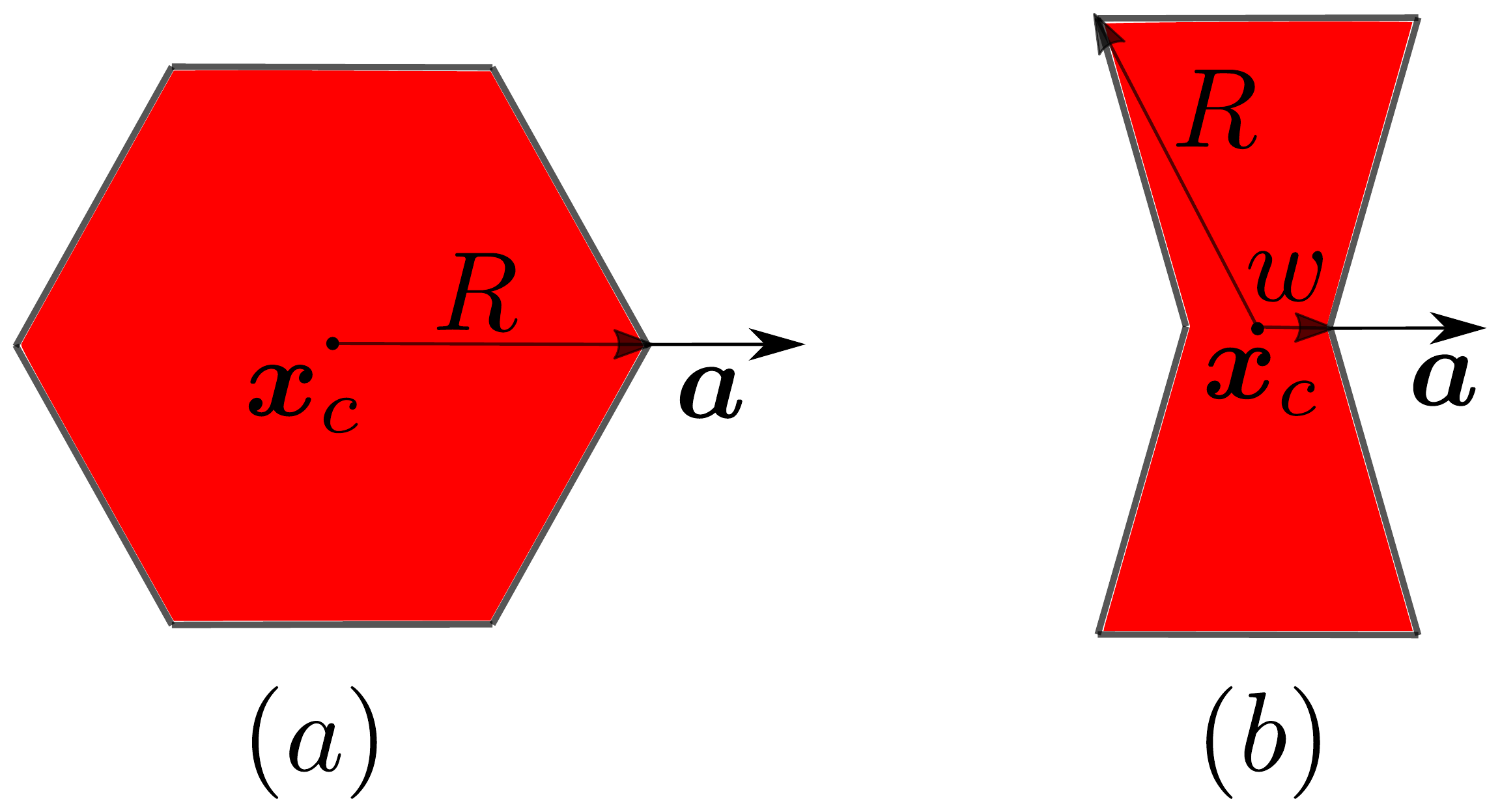}
  \caption{Examples of non-circular particle. {\bf (a)} Hexagon which can be generated using the three parameters: unit axis vector $\ba$, center $\bx_c$, and radius $R$. {\bf (b)} Concave polygon which can be generated using four parameters: unit axis vector $\ba$, radius $R$, half neck-width $w$, and center $\bx_c$.}\label{fig:noncircular}  
\end{figure}

In this section, we repeat few tests in previous sections using non-circular particles. We consider hexagon and concave polygon; see \autoref{fig:noncircular} for geometrical details. 

\begin{figure}[!htb] 
  \centering
  \includegraphics[width=0.8\textwidth]{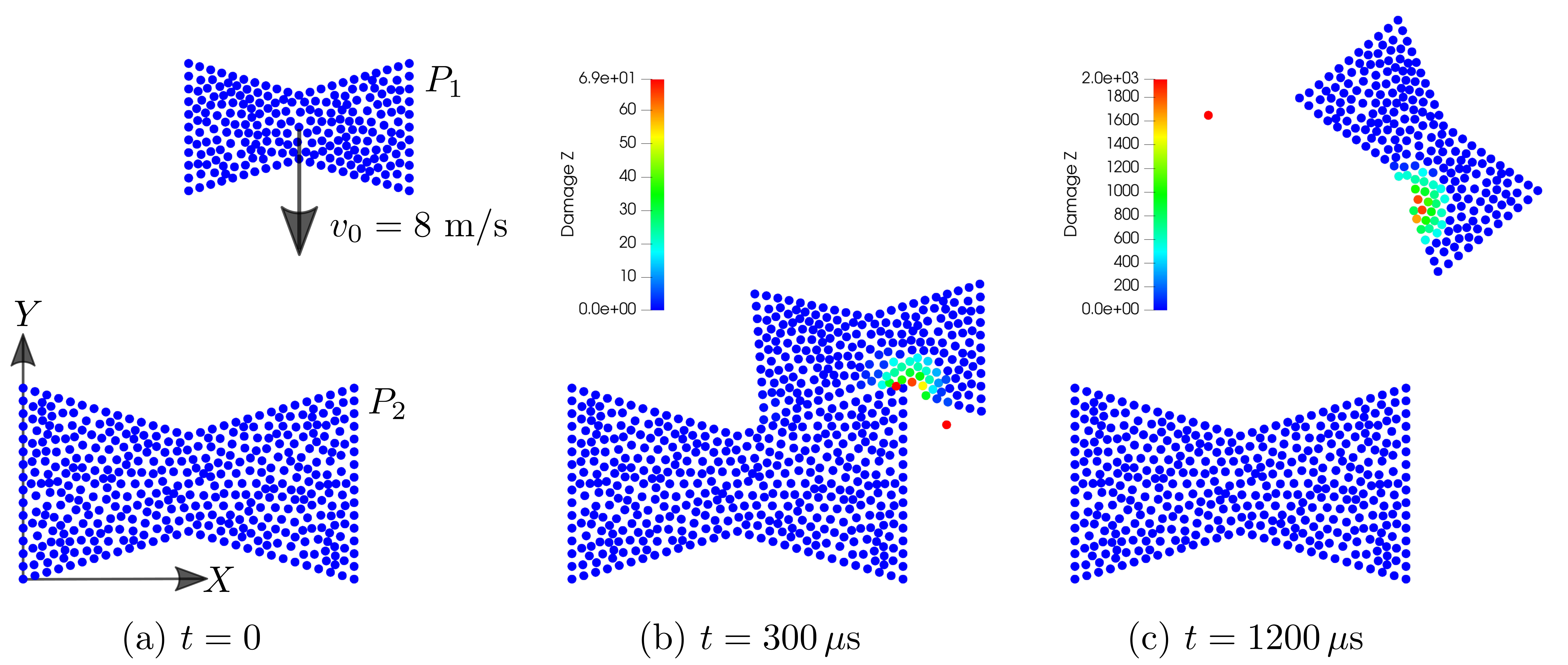}
  \caption{Non-circular two-particle fracture test. {\bf (a)} Initial configuration. Here, particle $P_1$ is generated from the parameters $R = 1$ mm, $\ba = (0,1)$, $w = 0.5*R$, and $\bx_c = (2.366,  4.3)$. And particle $P_2$ is generated from the parameters $R = 1.5$ mm, $\ba = (0,1)$, $w = 0.5*R$, and $\bx_c = (1.5,  1.5)$. Coordinates are in units of mm. {\bf (b)} and {\bf (c)} show the configuration and damage at two times.}\label{fig:twoPNonCircDamage}  
\end{figure}

\begin{figure}[!htb] 
  \centering
  \includegraphics[width=0.8\textwidth]{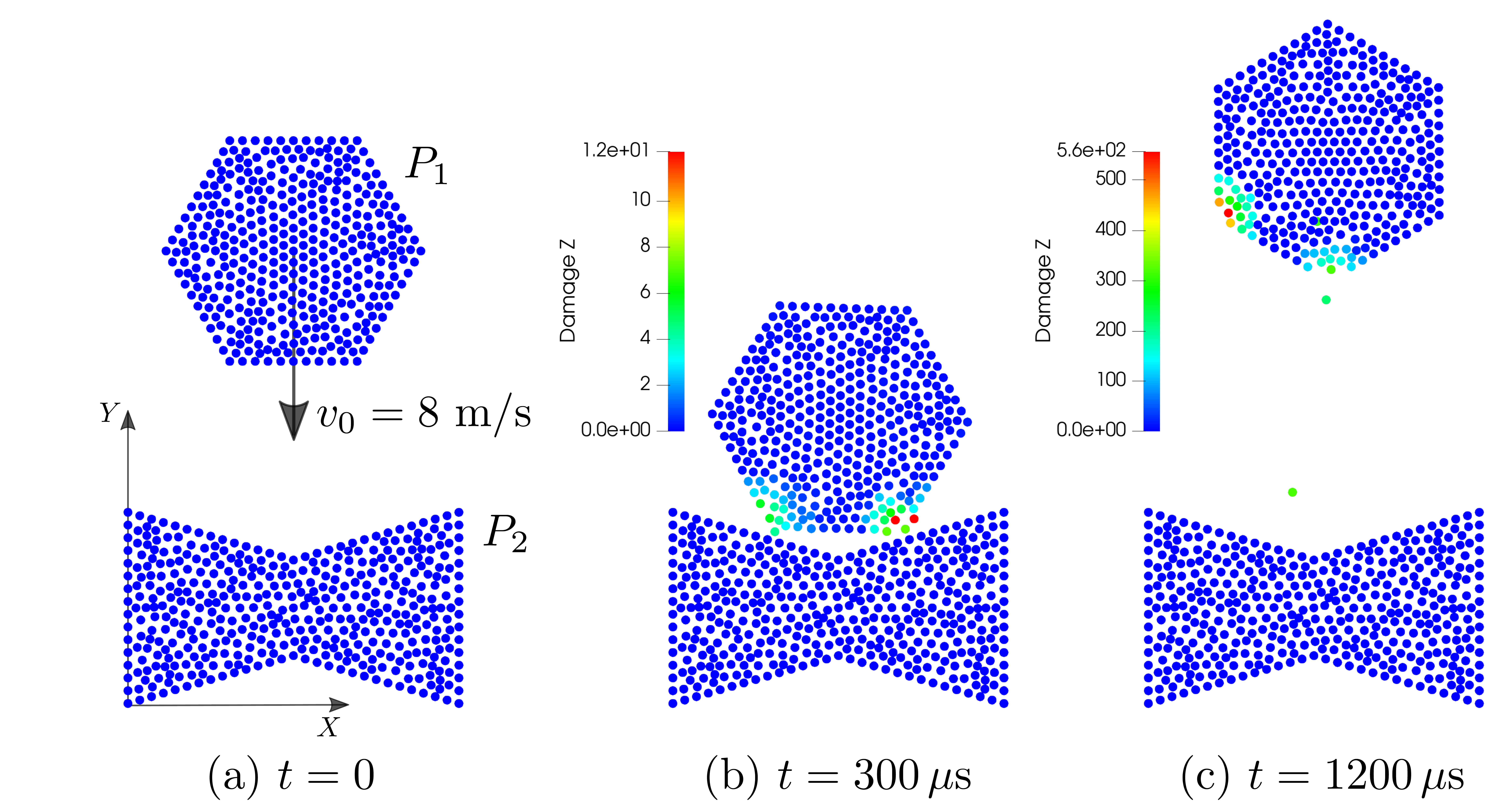}
  \caption{Non-circular two-particle fracture test. {\bf (a)} Initial configuration. Here, particle $P_1$ is generated from the parameters $R = 1$ mm, $\ba = (1,0)$, and $\bx_c = (1.5,  4.3)$. And particle $P_2$ is generated from the parameters $R = 1.5$ mm, $\ba = (0,1)$, $w = 0.5*R$, and $\bx_c = (1.5,  1.5)$. Coordinates are in units of mm. {\bf (b)} and {\bf (c)} show the configuration and damage at two times.}\label{fig:twoPNonCircHexDamage}  
\end{figure}

First, we repeat the two-particle fracture test in \autoref{ss:twop} where we replace the circular particles in \autoref{fig:twoPSetup} with concave particle shown in \autoref{fig:noncircular}. We assign top particle a downward velocity of $8$ m/s. We consider final time $T = 1200 \,\mu$s, time step  $\Delta t = 0.05 \, \mu$s, mesh size $h = 0.058$ mm, and horizon $\epsilon = 0.3$ mm. Both particles share the material properties M1. Damping acts at the centers of particles and the associated parameters are $\bar{\varepsilon}_n = 0.95$. Frictional force, similar to earlier cases, is switched off. 
In \autoref{fig:twoPNonCircDamage}, we show the initial configuration and damage. Next, we repeat the above test, replacing the particle on top with a hexagon-shaped particle. \autoref{fig:twoPNonCircHexDamage} shows the initial configuration and damage within the material at two times. 

\begin{figure}[!htb] 
  \centering
  \includegraphics[width=0.8\textwidth]{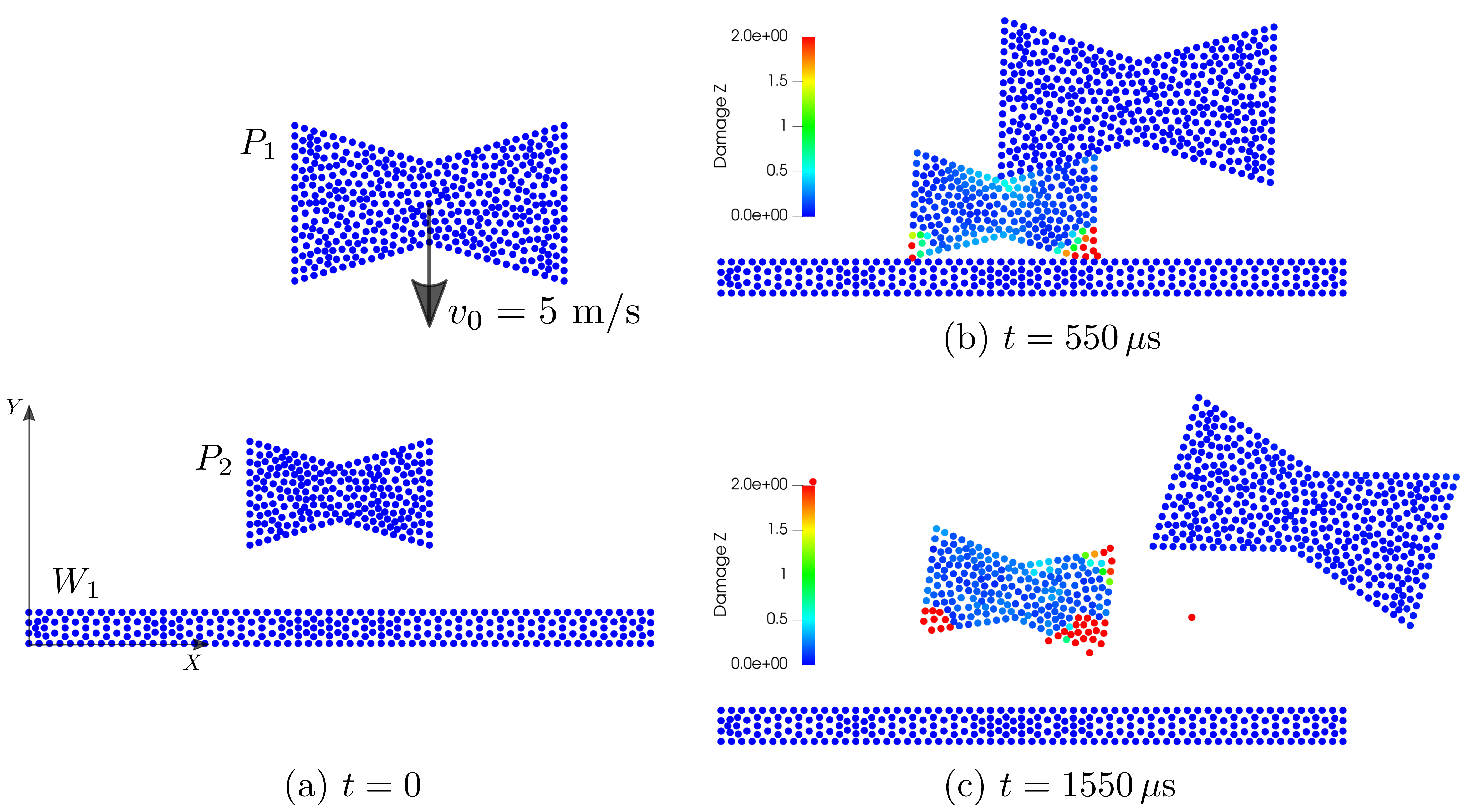}
  \caption{Non-circular two-particle with wall fracture test. {\bf (a)} Initial configuration. Here, particle $P_1$ is generated from the parameters $R = 1.5$ mm, $\ba = (0,1)$, $w = 0.5*R$, and $\bx_c = (1.866,  3.8)$. And particle $P_2$ is generated from the parameters $R = 1$ mm, $\ba = (0,1)$, $w = 0.5*R$, and $\bx_c = (1,  1)$. The rectangular wall $W_1$ is defined by the two corner points $(-2, -0.45)$ (left-bottom) and $(2, -0.15)$ (top-right). Coordinates are in units of mm. {\bf (b)} and {\bf (c)} show the configuration and damage at two times. We have fixed the upper damage to $2$ to show that the damage is also seen in the top-right corner region of the bottom particle.}\label{fig:twoPWallNonCircDamage}  
\end{figure}

As a last simulation in this section, we consider the two-particle with wall test where we replace the circular particles in \autoref{fig:twoPWSetup} with concave particle shown in \autoref{fig:noncircular}. The initial setup is shown in \autoref{fig:twoPWallNonCircDamage}(a). We take final time $T = 1600\,\mu$s and time step $\Delta t = 0.05\, \mu$s. We consider mesh size $h = 0.058$ mm, horizon $\epsilon = 3$ mm, damping parameter $\bar{\varepsilon} = 0.95$. Material properties of particles and wall are same and given by M1. The plot of damage at two times is shown in \autoref{fig:twoPWallNonCircDamage}(b,c).

\FloatBarrier

\subsection{Multi-particle compressive test} \label{ss:compressiveTest}
So far, we have shown the application of the model to settings involving two particles. Two-particle settings provide useful information and allow one to calibrate the contact parameters for the desired effect. Further, these serve the purpose of code validation and verification. Our results show that the model can be calibrated for damping effects and behaves consistently with varying parameters such as particle radius and material properties. Thus far, the applications also highlight the model's features to seamlessly capture inter-particle dynamics and intra-particle damage that may eventually result in total breakage under certain conditions. 

\begin{figure}[!htb] 
  \centering
  \includegraphics[width=\textwidth]{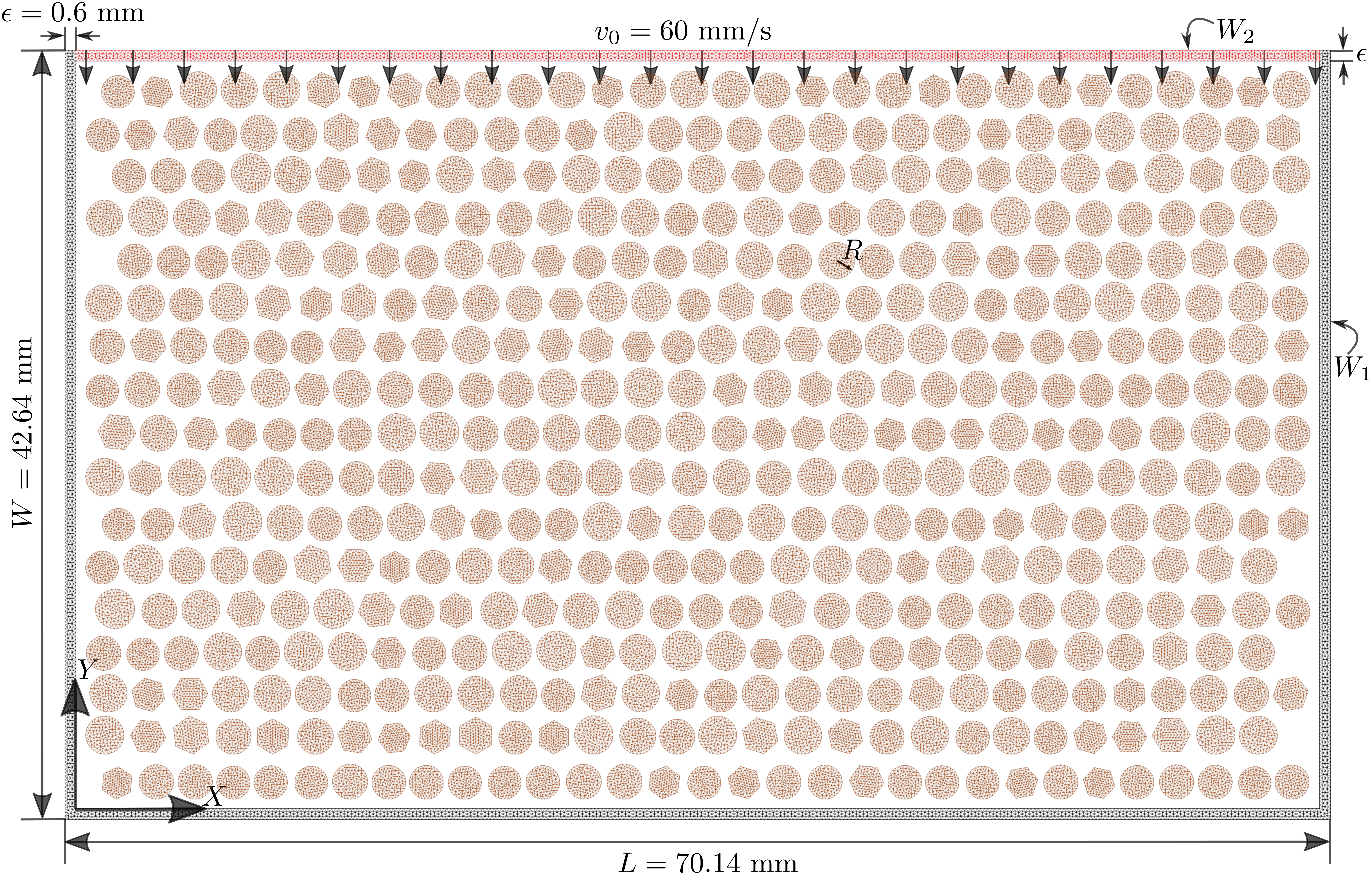}
  \caption{Multi-particle compressive test setup. Wall $W_2$ on top moves downwards with constant velocity $v_0$ whereas $W_1$ comprising of vertical walls and bottom wall is fixed. Particles are subjected to gravity of $g = 10$ m/s$^2$ downwards. We consider a random mixture of circular and hexagon-shaped particles. The radius of particles follow the uniform distribution as follows: $R \sim 1 + \mathcal{U}(-0.1, 0.1)$ (in units of mm). Particles are also given random rotation about their centroid. Centers of particles are arranged in a uniform grid. To not let particle centers aligned vertically, we randomly perturb the particles in the x-direction. The walls are of thickness same as horizon $\epsilon$ in peridynamics.}\label{fig:compresiveSetup}  
\end{figure}

In this section, we consider a slightly more complex setting involving 502 particles of varying radii in a rectangular container, see \autoref{fig:compresiveSetup}. We consider a random mixture of circular and hexagon-shaped particles. The top wall of the container is moving downwards at a constant velocity. Media is subjected to the downward gravity $g=10$ m/s$^2$. Particle radii are based on a distribution $R\sim 1 + \mathcal{U}(-0.1, 0.1)$ where units are in mm and $\mathcal{U}(a,b)$ denotes the uniform distribution with samples taking value between $a$ and $b$. Further, the particles are randomly perturbed in x-direction a little so that the particles' centers are not aligned vertically. Each particle is randomly rotated about its centers. \autoref{fig:compresiveSetup} shows other geometric details and the velocity of the top wall. Walls and particles are discretized using the Gmsh library, and later the mesh is converted to get the meshless discretization following \autoref{ss:peridemImp}. The minimum mesh size after discretization is $h = 0.116$ mm, and the contact radius is $0.95$ times the minimum mesh size. We first simulate the media for $T = 0.06$ s with time step $\Delta t = 0.1$ $\mu$s; during this simulation the particles settle down due to gravity. We then consider the current configuration at the end of the first simulation as the initial configuration and simulate additional $0.09$ s with the same time step (the total simulation time is $0.15$ s). In the second run, to bring the top wall closer to the particles, we modify the initial location of the top wall; after this, the initial position of the top wall's bottom edge is $0.0312$ m in the second run. 

\begin{figure}[!htb] 
  \centering
  \includegraphics[width=\textwidth]{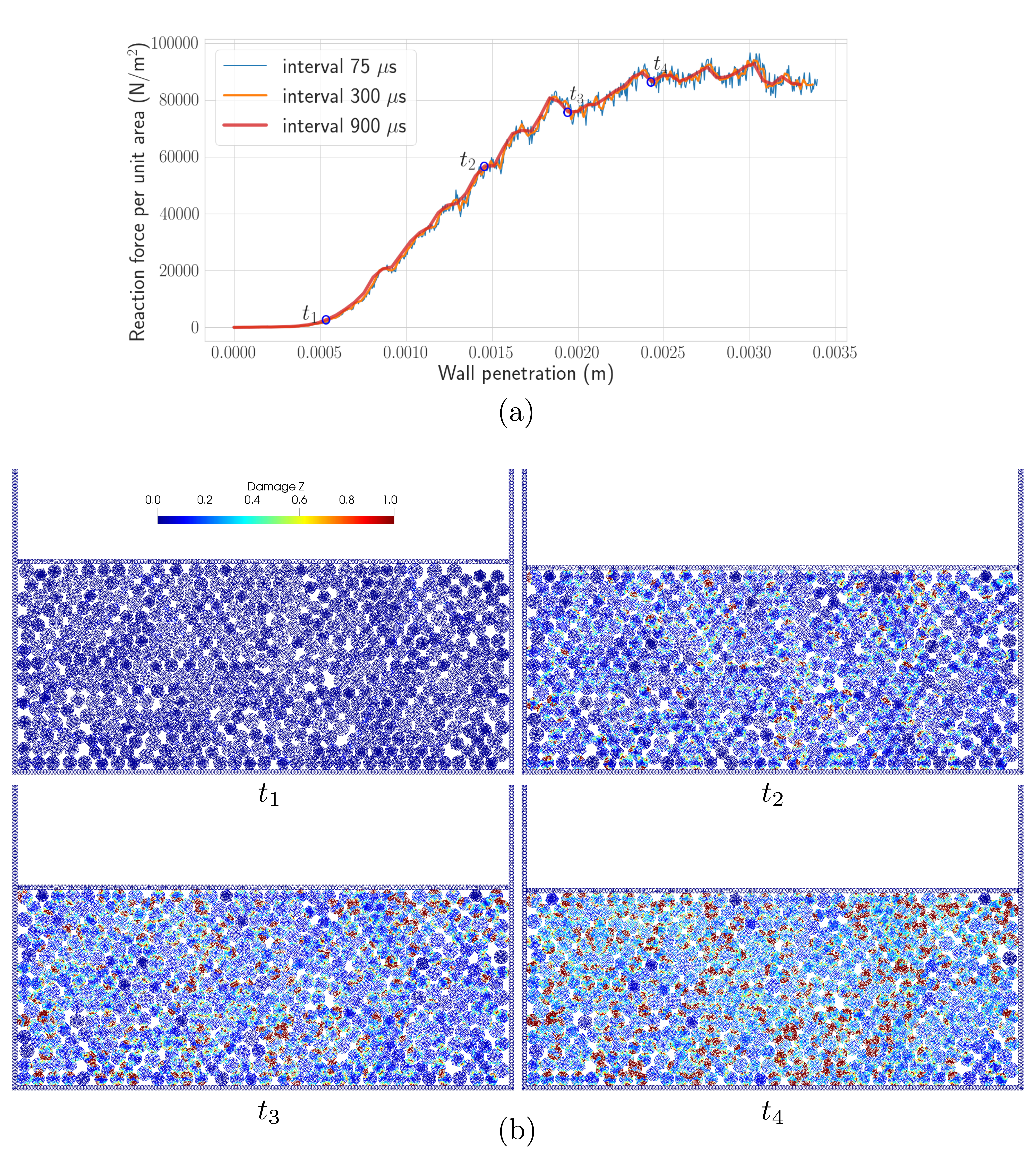}
  \caption{{\bf Top}: Plot of downward wall distance vs. reaction force (vertical component) per unit area. We average the force using the three different time intervals. This smoothes out the fluctuations in force due to the dynamic nature of the simulation. We show four marked time points $\{0.102, 0.118, 0.126, 0.134\}$ (in units of second) in the curve. Up to about $t_3$, the media's response is elastic to the increasing compressive loading; this changes near $t_3$, and the media starts to yield due to the softening of the particles forming the force chain. The media exhibits some strain hardening from $t_3$ to $t_4$, and beyond $t_4$, it displays a plastic failure due to an increased number of damaged bonds in particles. {\bf Bottom}: Configuration of particles and associated damage plot at four marked time points $\{0.102, 0.118, 0.126, 0.134\}$ (in units of second) in the top curve. Significant particle damage is visible at $t_4$.}\label{fig:compTestRF}  
\end{figure}

In \autoref{fig:compTestRF}(a), we plot the total reaction force (vertical component) per unit area on the moving wall with respect to the wall penetration. We identify 4 points with times $t_1 = 0.102, t_2 = 0.118, t_3 = 0.126, t_4 = 0.134$ (in units of second) on the force curve and plot the configuration of particles with damage in \autoref{fig:compTestRF}(b). The media starts experiencing the compressing action at time $t_1$. From $t_1$ to $t_2$, the media exhibits an elastic behavior with the force on wall increasing linearly with time. We see that at time $t_3$, the media yields a little due to the softening of the particles forming the force chain; see figure for $t_1$ in \autoref{fig:compTestRF}(b). The media exhibits some strain hardening from $t_3$ to $t_4$. From $t_4$, the media exhibits a plastic failure. The proposed model has potential application in the estimation of the effective strength of the particulate media.

\FloatBarrier

\section{Discussion and conclusion}\label{s:conclusion}
We have presented a new hybrid model that combines the advantages of the discrete element method (DEM) and peridynamics for more accurate simulations of the granular media. Numerical results show that the model is reliable under different scenarios, and parameters can be tuned to have the desired damping effect and contact stiffness. Under small deformation, the model behaves like an elastic body. However, situations such as high-velocity impacts, compressive loading from the surrounding walls, \etc, can cause significant damage and attrition in the particles and ultimately result in particle breakage. When spherical or polyhedral particles break, they no longer maintain a convex geometry, and the resulting inter-particle locking may become essential to capture particle dynamics accurately. The proposed model can seamlessly handle this scenario as the contact forces do not explicitly depend on particles' surface geometry since the contact is applied at the material point level. The model also correctly simulates the coefficient of restitution (CoR) in a two-particle impact test. A multi-particle compressive test shows the utility of the model to simulate particle damage and its progression. 

\begin{figure}[!htb]
  \centering
  \includegraphics[width=0.4\textwidth]{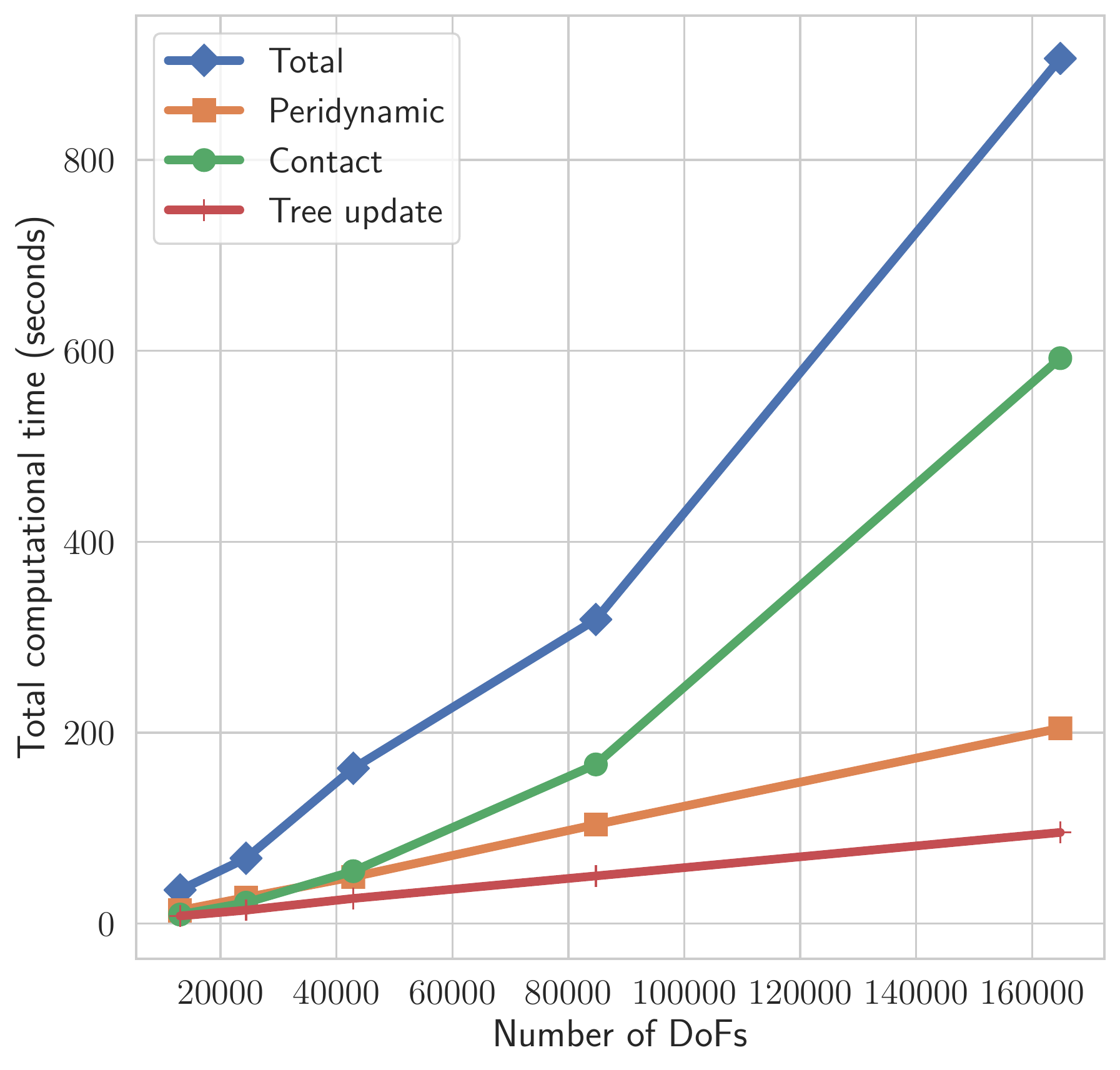}
  \caption{Total computational time for five tests with particles $25, 51, 96, 200, 403$. The number of degree of freedoms for these five tests are $13029, 24405, 42900, 84744, 164859$.}\label{fig:speedup}  
\end{figure}

With the use of PCL library \citet{Rusu_ICRA2011_PCL, muja2009fast}, we have been able to speed up the computation by the magnitude of orders and can simulate as many as 1000s of particles in a reasonable amount of time; the speed up gets better and better as the number of particles (or total degree of freedoms) increases. To shed some light on computational time, we performed the compressive test in \autoref{ss:compressiveTest} using a varying number of particles: we considered five tests with particles $25, 51, 96, 200, 403$. The total degree of freedoms (twice the number of discretized nodes in 2d) corresponding to these tests are $13029, 24405, 42900, 84744, 164859$ respectively. In \autoref{fig:speedup}, we plot the total computational time associated with different model components. We simulated $10000$ time steps with final time $T = 0.001$ seconds and time step $\Delta t = 0.1\,\mu$s. In all tests, we have utilized 12 threads. The computational time is almost linear thanks to the efficient neighbor search library. As the size of discretized node increases, we see an increase in contact computation compared to the peridynamics. The additional cost in contact is due to neighbor list calculation every time step. For the peridynamics calculation, the neighbor list is built only once at the beginning of the simulation.

There is a great deal of scope for further optimization of the computation. For example, the neighbor list for contact force is computed every time step, and for this, the k-dimensional tree \citet{Rusu_ICRA2011_PCL, muja2009fast} are rebuilt using the updated location of discretized nodes. The cost of such an operation is not high; however, for conditions where particle motion is not rapid, it will be more efficient to build the tree and contact neighbor list every $n\geq 1$ time steps. The pairwise calculations are not computationally heavy and are large in number (for peridynamics and nonlocal contact). These simpler calculations suggest that GPU can perform these calculations much faster and in parallel, reducing the compute time. From a modeling point of view, as expected, a significant contribution to the computational cost is from peridynamics; this motivates us to consider local continuum mechanics models or even rigid body motion for particles with small deformation.

Additionally, the PeriDEM framework can benefit from the local-nonlocal coupling approach where the nonlocal calculations are restricted to a small region in the media. Another direction for further speed up and large-scale application is artificial neural networks (ANNs). ANNs can replace the peridynamics calculation and contact calculation allowing faster computation. Future works will explore some of the possibilities listed above. 

While this work's objective was to introduce a new model that can benefit many fields such as construction (cement and rock fragments), pharmaceutics (tablets), and transportation (ballast), all fields that rely on accurate modeling of powder or granular mechanics, particle wear, and breakage can benefit from this model. Future works will be towards the application of the model to specific challenging problems. While the model proposed here is purely mechanical, it is possible to introduce new effects such as a change in strength due to temperature, fluid-structure coupling, \etc.


\end{document}